\documentclass[nofootinbib,preprintnumbers,prd,superscriptaddress,showpacs,twocolumn]{revtex4}

\usepackage{amsmath}
\usepackage[titletoc]{appendix}
\usepackage{color}
\usepackage{hyperref}

\usepackage[rightcaption]{sidecap}
\usepackage{subfigure}
\usepackage{comment}

\usepackage{graphicx}
\usepackage{amsmath}
\usepackage{amsfonts}
\usepackage{amsthm}
\usepackage{mathrsfs}
\usepackage{amssymb}
\usepackage{epsfig}
\usepackage{amscd}
\usepackage{dcolumn}
\usepackage{bm}
\usepackage{natbib}
\usepackage{url}
\usepackage{xspace}

\usepackage{dcolumn}

\usepackage{array}
\usepackage{ctable}
\usepackage{multirow}
\usepackage{siunitx}
\usepackage{longtable}
\usepackage{tabularx}
\usepackage{booktabs}

\usepackage{amsthm}
\usepackage{mathrsfs}

\usepackage{epsfig}
\usepackage{amscd}

\usepackage{bm}
\usepackage{natbib}
\usepackage{url}
\usepackage{xspace}
\usepackage{kantlipsum}

\graphicspath{{Graphics/}}

\def\be{\begin{equation}}
\def\ee{\end{equation}}
\def\bea{\begin{eqnarray}}
\def\eea{\end{eqnarray}}

\definecolor{vividviolet}{rgb}{0.62, 0.0, 1.0}
\definecolor{amaranth}{rgb}{0.9, 0.17, 0.31}
\definecolor{palatinateblue}{rgb}{0.15, 0.23, 0.89}
\definecolor{brightpink}{rgb}{1.0, 0.0, 0.5}
\definecolor{cornflowerblue}{rgb}{0.39, 0.58, 0.93}
\definecolor{deepcarminepink}{rgb}{0.94, 0.19, 0.22}
\definecolor{radicalred}{rgb}{1.0, 0.21, 0.37}

\hypersetup{ linktoc=all,
    colorlinks, linkcolor={palatinateblue},
    citecolor={brightpink}, urlcolor={amaranth}
}

\begin{document}

\title{A barotropic alternative to Early Dark Energy for alleviating the $H_0$ tension}

\author{Youri Carloni}
\email{youri.carloni@unicam.it}
\affiliation{Universit\`a di Camerino, Via Madonna delle Carceri, Camerino, 62032, Italy.}
\affiliation{INAF - Osservatorio Astronomico di Brera, Milano, Italy.}
\affiliation{Istituto Nazionale di Fisica Nucleare, Sezione di Perugia, Perugia, 06123, Italy.}

\author{Orlando Luongo}
\email{orlando.luongo@unicam.it}
\affiliation{Universit\`a di Camerino, Via Madonna delle Carceri, Camerino, 62032, Italy.}
\affiliation{INAF - Osservatorio Astronomico di Brera, Milano, Italy.}
\affiliation{Department of Nanoscale Science and Engineering, University at Albany SUNY, Albany, NY 12222, USA.}
\affiliation{Istituto Nazionale di Fisica Nucleare, Sezione di Perugia, Perugia, 06123, Italy.}
\affiliation{Al-Farabi Kazakh National University, Al-Farabi av. 71, 050040 Almaty, Kazakhstan.}

\begin{abstract}
We propose a cosmological scenario in which, beyond matter and radiation, an additional barotropic fluid with positive equation of state $\omega_s$ contributes to the cosmic energy budget, in contrast to Early Dark Energy (EDE). We investigate the theoretical implications of this framework, here dubbed the $\Lambda_{\omega_s}$CDM model, at both the background and perturbative levels, exploring its impact on the expansion history and structure formation. We show that, while remaining subdominant at late times and therefore consistent with current observational bounds, the additional fluid modifies the early-time expansion rate, leading to a higher inferred value of the Hubble constant. Thus, we perform a full Bayesian analysis using a modified version of the \texttt{CLASS} Boltzmann code interfaced with \texttt{MontePython}, considering combinations of \textit{Planck} 2018 Cosmic Microwave Background (CMB) data, DESI DR2 Baryon Acoustic Oscillations (BAO) measurements, Pantheon Type Ia supernovae (SNe Ia), and SH0ES determinations of $H_0$. We find that the inclusion of the SH0ES prior, $H_0 = 73.04 \pm 1.04\,\mathrm{km/s/Mpc}$, leads to a preference for a nonvanishing barotropic fluid. In particular, we obtain $\omega_s = 0.290^{+0.017(0.021)}_{-0.007(0.028)}$ and density $10^5\Omega_s = 1.47^{+0.35(1.14)}_{-0.62(0.94)}$ for the dataset combination CMB + BAO + Pantheon + SH0ES, and $\omega_s = 0.302^{+0.024(0.034)}_{-0.013(0.038)}$ and $10^5\Omega_s = 1.21^{+0.31(1.10)}_{-0.65(0.86)}$ when BAO data are excluded. We further compare our scenario with the EDE framework and show that, statistically, no strong evidence is found against the $\Lambda_{\omega_s}$CDM model. Finally, we provide a physical interpretation of our fluid in terms of matter with pressure, indicating that the standard cosmological model may be incomplete in its current minimal formulation.
\end{abstract}


\maketitle
\tableofcontents


\section{Introduction}

The standard cosmological model is based on the existence of fundamental barotropic fluids, such as matter, radiation, neutrinos, and so on \cite{Mukhanov:1992me,Hu:1995hf,Hu:2001bc,Lesgourgues:2006nd}. However, after the discovery of cosmic speed up, one more constituent has been proposed, namely the cosmological constant that, in its original formulation, originates from primordial quantum fluctuations \cite{Weinberg:1988cp,Peebles:2002gy,Padmanabhan:2002ji}. 

All the attempts to describe the cosmological acceleration with an evolving dark energy contribution faced conceptual and observational inconsistencies, suggesting that the standard background paradigm, namely the $\Lambda$CDM model, has proven to be a robust and viable framework to describe the late- and early-scale structures of the Universe.

However, very recently cosmological tensions have reopened the intrinsic issues\footnote{The existence of a cosmological constant naturally leads to the fine-tuning and coincidence issues \cite{Ratra:1987rm,Doran:2001rw,Martin:2012bt}, which are so far unsolved within the standard cosmological framework; analogously, any dark energy model or alternative to Einstein's gravity does not usually justify why the cosmological constant is not necessary, thus leaving the cosmological constant problem still the object of strong debate in theoretical physics \cite{Garriga:2000cv,Tsujikawa:2010sc,Sola:2013gha}.} associated with the $\Lambda$CDM model \cite{Verde:2019ivm,DiValentino:2020zio,DiValentino:2020vvd,Schoneberg:2021qvd, Poulin:2022sgp,Abdalla:2022yfr,CosmoVerseNetwork:2025alb}, while the results of BOSS mission first \cite{BOSS:2012dmf} and, above all, DESI collaboration later \cite{DESI:2024mwx,DESI:2025zgx} provided evidence for a possible dynamical dark energy model to describe the large-scale dynamics of the Universe. 

To solve the cosmological tension on $H_0$, several models have been proposed, among which, probably the easiest, consists of introducing one more scalar field, {\it i.e.}, the EDE constituent \cite{Poulin:2018dzj, Kamionkowski:2022pkx, Poulin:2023lkg}. This was initially thought to be always present throughout the entire cosmic evolution, while later suggested as a contribution at primordial times. 

The underlying mechanism is straightforward, {\it i.e.}, EDE models provide a phenomenological mechanism to alleviate the Hubble tension by increasing the pre-recombination expansion rate and reducing the sound horizon. However, this mechanism typically relies on a highly tuned scalar field that becomes dynamically relevant only within a narrow redshift window around recombination and rapidly dilutes afterwards. 

In particular, EDE models introduce additional parameters whose constraints can depend significantly on the dataset combination and on the statistical analysis adopted. While CMB data alone generally do not provide compelling evidence for EDE, the inclusion of external data, especially local determinations of the Hubble constant such as SH0ES, can increase the preference for a nonvanishing EDE component \cite{Poulin:2018cxd,Hill:2020osr}. 

At the same time, EDE can shift other cosmological observables, and its inferred viability is sensitive to the treatment of Large Scale Structure (LSS) data, prior choices, and marginalization effects \cite{Ivanov:2020ril,Smith:2020rxx,Herold:2021ksg}. 

Accordingly, some analyses find that high-precision CMB and LSS data place significant constraints on the EDE parameter space \cite{Hill:2020osr,Ivanov:2020ril}, whereas others conclude that EDE can still yield fits statistically comparable to $\Lambda$CDM once different priors or statistical approaches are considered \cite{Smith:2020rxx,Herold:2021ksg}.
Therefore, current observations do not yet provide a fully settled assessment of EDE, and its status as a possible resolution of the $H_0$ tension remains under investigation.

Hence, EDE represents an effective but somewhat \emph{ad hoc} solution, whose success is largely driven by its phenomenological flexibility, although its underlying motivation finds support within the string axiverse framework \cite{Svrcek:2006yi,Arvanitaki:2009fg,Marsh:2015xka,Visinelli:2018utg}. Nevertheless, while several realizations based on scalar fields have been proposed \cite{Poulin:2018cxd,Herold:2021ksg,Bella:2026zuk}, a clear and compelling embedding within a fundamental theory is still lacking.

\emph{This raises the question of whether a simpler and more generic mechanism can replace EDE along the entire Universe evolution}. 

In Ref. \cite{Carloni:2025jlk}, it has been pioneering proposed that one more barotropic fluid can be present in the cosmological realm. The fluid satisfies the Zeldovich limit \cite{Zeldovich:1961sbr}, being simply subdominant with respect to matter and radiation in the majority of cosmic evolution. 

In this work, motivated by the above findings, we propose how cosmology may be rewritten in terms of this new fluid, fixing theoretical limits and numerical constraints on its cosmological presence. To do so, we first ensure that its equation of state $\omega_s$ relies on the Zeldovich limit, appearing positive and very different from dark energy. In this respect, we reformulate the Friedmann equations and show how the corresponding solutions on the scale factor, Hubble rate and distances are modified as a consequence of such a further constituent. We justify its existence by remarking that the fluid sub-dominates with respect to matter and radiation after the CMB, and show when we expect equivalence periods between our new constituent and the standard cosmological fluids. We then show that late- and early-times are only partially affected by the fluid itself, explaining possible physical reasoning that has led observations not to find this extra-constituent so far. A direct comparison of our new fluid with EDE is thus performed, with a hierarchical choice of data sets, including and excluding BAO data\footnote{The use of BAO as a standard ruler is intrinsically limited by late-time non-linear effects, which smear and shift the acoustic peak in a redshift- and model-dependent way, thereby requiring cosmology-dependent template fitting and reconstruction procedures that may introduce systematic biases and underestimate distance uncertainties \cite{Anselmi:2018vjz,ODwyer:2019rvi,Anselmi:2022exn}.}. To do so, we employ \textit{Planck} 2018 CMB data, the DESI DR2 BAO dataset, the Pantheon and SH0ES catalogs, as well as the Observational Hubble data (OHD). We perform a Monte Carlo Markov Chain (MCMC) analysis using a modified version of the publicly available CLASS code \cite{Blas:2011rf}. At the perturbative level, we explore the matter growth perturbations and the impact of the new fluid on structure formation. Our outcomes are thus intertwined and we end up with how physically our fluid may be interpreted in view of cosmological modern probes. We thus conclude that \emph{the existence of one more fluid in cosmological realms is possible}, whose interpretation is due to a quasi-relativistic component, reinterpreted in terms of a \emph{matter with pressure new term}, whose equation of state resembles radiation, albeit being quite different from it. In this respect, we show that deforming mass and radiation is unlikely to obtain such an extra contribution. Accordingly, we propose that this treatment, conventionally dubbed $\Lambda_{\omega_s}$CDM paradigm, offers a minimal modification of the standard background model and substitutes the need of a primordial EDE with an extra barotropic fluid that alleviates the $H_0$ tension. 

The manuscript is organized as follows. In Sec. \ref{Sec.2}, we describe the idea of introducing a novel fluid. Thus, we explore the background and early-time consequences of this recipe in Sec. \ref{late} and Sec. \ref{early}, respectively. In Sec. \ref{Sec.3}, we perform an MCMC analysis by combining different probes from early to late times in order to test our model. In Sec. \ref{sec6}, we carry out a statistical comparison between $\Lambda_{\omega_s}$CDM and EDE by considering CMB \textit{Planck} 2018, DESI DR2 BAO, Pantheon, and the SH0ES prior. In Sec. \ref{sec7}, we provide a physical interpretation of our new fluid, excluding its possible origin in terms of matter and/or radiation deformations. Finally, in Sec. \ref{Sec.8}, we present our conclusions and perspectives.


\section{Cosmology with an additional barotropic fluid}\label{Sec.2}

The so-called $\Lambda_{\omega_s}$CDM model, introduced in Ref.~\cite{Carloni:2025jlk}, extends the concordance paradigm by adding a new barotropic fluid, interpreting such a new component in terms of \emph{matter with pressure}, which contributes non-negligibly at early-times, while being subdominant at late times. 

The additional fluid component fulfills the Zeldovich limit, providing a positive equation of state, $0<\omega_s<1$, being different from pure dust and stiff matter, $\omega_s\neq0$ and $\omega_s\neq1$, respectively. The density of the fluid, in terms of the scale factor $a$, then formally reads,
\begin{equation}\label{eq:rhos_standard}
    \rho_s(a)\equiv \rho_{s}a^{-3(1+\omega_s)},
\end{equation}
with $\omega_s$ constant. This is quite different from dark energy, which instead ensures a negative equation of state, plus the additional requirement of being subdominant with respect to both matter and radiation throughout the entire cosmic expansion history after the CMB \cite{Carloni:2025jlk}. For the sake of completeness, from Fig. \ref{fig:Omegatot}, the fluid appears to dominate over matter at redshifts compatible with the radiation-dominated hot plasma epoch. A naive consequence within these two theoretical scenarios will be object of future efforts. 

In principle, this is not the first occurrence of incorporating additional fluids into the Hubble parameter, motivating the existence of a further constituent. Besides the resounding case of dark energy \cite{Copeland:2006wr}, another remarkable possibility is to ensure that cosmology is driven by stiff matter; see Ref. \cite{Chavanis:2014lra}. However, this fluid, appearing ultra-relativistic, is disfavored by observations \cite{Dutta:2010cu,Duval:2024jsg}, while being contemplated inside Eq. \eqref{eq:rhos_standard}, which does not fix any bounds on the free parameters, $\omega_s$ and $\rho_s$. The fluid is therefore not useful for easing the Hubble tension.

\section{Background consequences of our novel fluid}\label{late}

Cosmology in the presence of one more fluid, different from pressureless matter and radiation, clearly has consequences on the background. This leads to modifying the Friedmann equations, providing modifications of the standard Hubble parameter. Here, we explore the cases in which our fluid, defined in Eq. \eqref{eq:rhos_standard}, influences the overall evolution at the level of background, showing its impact on radiation and dust, respectively. Further, we underline that our new ingredient might be subdominant to dust and radiation after the CMB, implying no direct observational modification of our Universe at late times. To this end, we also compare it with a cosmological constant term. Our goal is not to alter the late-time behavior of the Hubble expansion rate but rather to affect the early-time dynamics, providing a viable resolution to the Hubble tension.

\subsection{The Friedmann equations}

In the $\Lambda_{\omega_s}$CDM cosmology, we proposed a Universe composed of matter with pressure, radiation, dust matter, and dark energy, treated as non-interacting components. 
We assume a spatially flat geometry, in agreement with CMB measurements.

Therefore, the total dimensionless energy density is obtained by summing the contributions of each component as

\begin{equation}
    \Omega(a)=\frac{\Omega_s}{a^{3(1+\omega_s)}}+\frac{\Omega_r}{a^4}+\frac{\Omega_m}{a^{3}}+\Omega_{\Lambda},\label{eq:tot}
\end{equation}

leading to the following Hubble parameter

\begin{equation}
    H(a)=H_0 \sqrt{\frac{\Omega_s}{a^{3(1+\omega_s)}}+\frac{\Omega_r}{a^{4}}+\frac{\Omega_m}{a^{3}}+\Omega_{\Lambda}},\label{eq:hubble}
\end{equation}
with $\Omega_s+\Omega_r+\Omega_m+\Omega_\Lambda=1$.

As a barotropic fluid, matter with pressure satisfies the Zeldovich limit, $\omega_s > 0$.

\begin{figure}
\includegraphics[width=0.5\textwidth,clip]{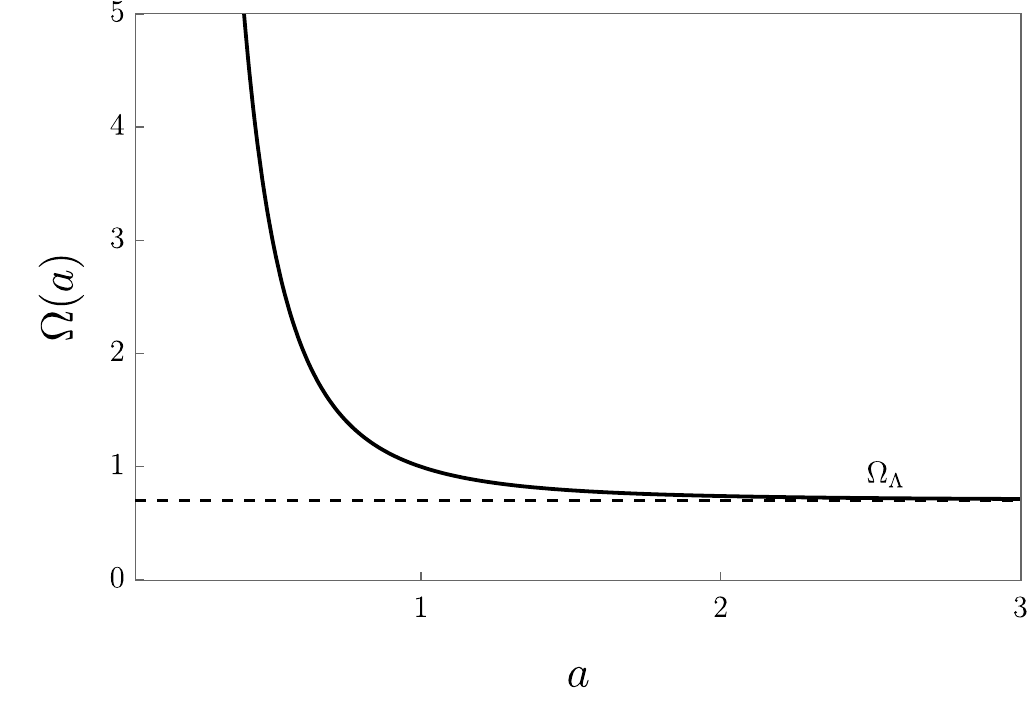}
\caption{Behavior of the total dimensionless energy density $\Omega(a)$ as a function of the scale factor $a$. 
In the limit $a \to 0$, $\Omega(a) \to \infty$, whereas for $a \to \infty$, $\Omega(a) \to \Omega_\Lambda$. For matter with pressure, we consider the values of Ref. \cite{Carloni:2025jlk}, {\it i.e.}, $\omega_s=0.294$, and $\Omega_s=1.62\times 10^{-5}$.We also take $\Omega_m = 0.296$ and $\Omega_r = 8.4 \times 10^{-5}$.}
\label{fig:Omegatot}
\end{figure}

\begin{figure}
\includegraphics[width=0.5\textwidth,clip]{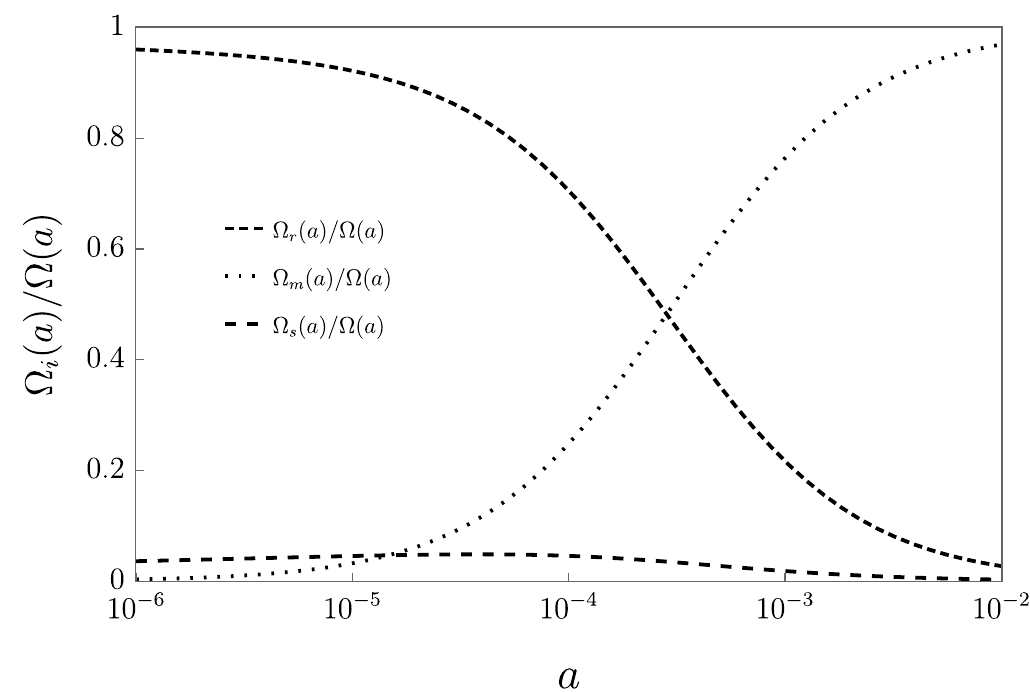}
\caption{Evolution of the dimensionless energy densities of each component as a function of the scale factor $a$ in the early-time regime, where the new fluid is more significant. 
Dark energy is neglected, as it is negligible at early-times. 
The parameter values are the same as in Fig.~\ref{fig:Omegatot}.}
\label{fig:Omegai}
\end{figure}

The total dimensionless energy density in Eq.~\eqref{eq:tot} is shown in Fig.~\ref{fig:Omegatot}, where the cosmological parameters are taken from our previous analysis in Ref.~\cite{Carloni:2025jlk}. 
We also consider the evolution of each component, $\Omega_i(a)/\Omega(a)$, in Fig.~\ref{fig:Omegai}, which shows that the contribution of matter with pressure emerges as $a \to 0$.

At this point, since $(H(a)/H_0)^2=\Omega(a)$, the evolution of the scale factor is provided by

\begin{equation}
\int_{a_i}^{a}
\frac{d\tilde{a}}{\tilde{a} \sqrt{
\frac{\Omega_{s}}{\tilde{a}^{3(1+\omega_s)}}
+
\frac{\Omega_{r}}{\tilde{a}^{4}}
+
\frac{\Omega_{m}}{\tilde{a}^{3}}
+
\Omega_{\Lambda}
}}
= H_0 t,\label{eq:a}
\end{equation}
where $a_i$ denotes the initial value of the scale factor.

From now on, we study the analytical solution of Eq.~\eqref{eq:a}, neglecting $\Omega_\Lambda$, which is subdominant in the early Universe, where our fluid is most relevant.

\subsubsection{Matter with pressure and radiation}

We analyze the early-time regime where radiation and matter with pressure dominate, neglecting dust matter. In this limit, Eq.~\eqref{eq:a} reduces to

\begin{equation}
\int_{a_i}^{a} \frac{\tilde{a}\,d\tilde{a}}{\sqrt{\Omega_r+\Omega_s \tilde{a}^{1-3\omega_s}}}=H_0t.
\end{equation}

\noindent
This integral can be evaluated analytically using the identity

\begin{equation}
\int \frac{\tilde{a}\,d\tilde{a}}{\sqrt{A + B \tilde{a}^{n}}}
=
\frac{\tilde{a}^2}{2\sqrt{A}}
\,{}_2F_1\!\left(
\frac{1}{2},\frac{2}{n};
1+\frac{2}{n};
-\frac{B}{A}\tilde{a}^{n}
\right),
\end{equation}
valid for $n \neq 0$.
Here, ${}_2F_1(a,b;c;z)$ denotes the Gauss hypergeometric function, defined as
\begin{equation}
{}_2F_1(a,b;c;d)=\sum_{k=0}^{\infty}
\frac{(a)_k (b)_k}{(c)_k}\frac{d^k}{k!},
\end{equation}
where $()_k$ represents the Pochhammer symbol. 

The series converges for $|d|<1$ and can be extended to other values of $d$ by analytic continuation.
In our case, we identify $
A=\Omega_r, \qquad B=\Omega_s, \qquad n=1-3\omega_s$, resulting in the following implicit analytical solution by choosing $a_i=0$
\begin{equation}
H_0 t=
\frac{a^2}{2\sqrt{\Omega_r}}
{}_2F_1\!\left(
\frac{1}{2},\frac{2}{1-3\omega_s};
1+\frac{2}{1-3\omega_s};
-\frac{\Omega_s}{\Omega_r}a^{1-3\omega_s}
\right).
\end{equation}
In general, it cannot be inverted analytically to obtain $a(t)$ in closed form.

The physical behavior of our novel fluid depends crucially on the value of the barotropic parameter $\omega_s$. We consider the following regimes.

\begin{itemize}

\item[-]$\omega_s < 1/3$ (radiation-dominated regime).

In this case, $1-3\omega_s > 0$, and therefore
\begin{equation}
a^{1-3\omega_s} \rightarrow 0 \quad \text{as} \quad a \rightarrow 0.
\end{equation}
Radiation dominates over matter with pressure, and the scale factor becomes

\begin{equation}
a(t) \simeq \left(2\sqrt{\Omega_r}H_0t\right)^{1/2},
\end{equation}

recovering the standard radiation-dominated expansion, with subdominant corrections due to the additional fluid.

\item[-] $\omega_s > 1/3$ (early-time fluid domination).

Here, we have $1-3\omega_s < 0$, determining 
\begin{equation}
a^{1-3\omega_s} \rightarrow \infty \quad \text{as} \quad a \rightarrow 0.
\end{equation}
The barotropic fluid dominates the expansion as

\begin{equation}
H^2 \simeq H_0^2 \frac{\Omega_s}{a^{3(1+\omega_s)}},
\end{equation}
leading to the following form of the scale factor

\begin{equation}
a(t) \simeq
\left[
\frac{3}{2}(1+\omega_s)\sqrt{\Omega_s}\,H_0 t
\right]^{\frac{2}{3(1+\omega_s)}}.
\end{equation}

Exploring cosmic bounds, through our statistical analysis, will shed light on the values acquired by $\omega_s$ that converge to $\omega_s<1/3$. Theoretically speaking, this is absolutely required at late-times, to avoid that our fluid dominates over other species. At early-times, this will be certified as well, requiring moreover the need of healing the $H_0$ tension, as we will consider later in the text.

\end{itemize}

\subsubsection{Matter with pressure versus dust}

We now consider a cosmological scenario in which the expansion is driven by dust matter and \emph{matter with pressure}, neglecting radiation. 

This is particularly important for our purposes since it permits us to characterize the fluid introduced in Eq. \eqref{eq:rhos_standard} as a relativistic matter-like species, different from exotic EDE models \cite{Bella:2026zuk}, exotic scalar fields \cite{Luongo:2018lgy,Belfiglio:2022qai} and/or axion-like contributions \cite{Vagnozzi:2023nrq}. In this case, we solve

\begin{equation}
\int_{a_i}^a \frac{\tilde{a}^{1/2}\,d\tilde{a}}{\sqrt{\Omega_m+\Omega_s \tilde{a}^{-3\omega_s}}}=H_0 t.
\end{equation}

The integral can be again evaluated analytically in terms of the Gauss hypergeometric function as

\begin{equation}
\int \frac{\tilde{a}^{1/2}\,d\tilde{a}}{\sqrt{A+B \tilde{a}^{p}}}
=
\frac{2}{3}\frac{\tilde{a}^{3/2}}{\sqrt{A}}
\,{}_2F_1\!\left(
\frac{1}{2},-\frac{1}{2\omega_s};
1-\frac{1}{2\omega_s};
-\frac{B}{A}\tilde{a}^{p}
\right),
\label{eq:dust_s_34}
\end{equation}

\noindent
where $A=\Omega_m$, $B=\Omega_s$, and $p=-3\omega_s$.

Therefore, by selecting $a_i=0$, the analytical solution is given here by

\begin{equation}
H_0 t=
\frac{2}{3}\frac{a^{3/2}}{\sqrt{\Omega_m}}
{}_2F_1\!\left(
\frac{1}{2},-\frac{1}{2\omega_s};
1-\frac{1}{2\omega_s};
-\frac{\Omega_s}{\Omega_m}
a^{-3\omega_s}
\right),
\end{equation}

which, also in this case, cannot be inverted to yield a closed form for $a(t)$.

Then, the cosmological dynamics depends on the competition between dust matter and matter with pressure as follows.

\begin{itemize}

\item[-]Early-time regime.

As $a \rightarrow 0$, we get $a^{-3\omega_s} \to \infty$, and matter with pressure dominates over dust. In this limit,

\begin{equation}
H^2 \simeq H_0^2 \frac{\Omega_s}{a^{3(1+\omega_s)}},
\end{equation}

leading to

\begin{equation}
a(t) \simeq
\left[
\frac{3}{2}(1+\omega_s)\sqrt{\Omega_s}\,H_0 t
\right]^{\frac{2}{3(1+\omega_s)}}.
\end{equation}

\item[-] Late-time regime.

Instead, for sufficiently large $a$, the dust component dominates, and we recover

\begin{equation}
a(t) \simeq \left(\frac{3}{2}\sqrt{\Omega_m}H_0 t\right)^{2/3}.
\end{equation}

\end{itemize}

In contrast to the radiation-dominated case, for any $\omega_s>0$, the presence of matter with pressure modifies the early-time dynamics, while the standard dust-dominated expansion is recovered at late times.

\subsubsection{Full three-fluid early-time cosmology}

In order to quantify the impact of the additional barotropic component, we compare the numerical solution of the early-time three-fluid cosmology with the exact analytical solution corresponding to a Universe composed only of radiation and dust.

In the standard two-fluid case, the Hubble parameter is given by
\begin{equation}
H(a) = H_0 \sqrt{\frac{\Omega_r}{a^4} + \frac{\Omega_m}{a^3}},
\end{equation}
and the evolution of the scale factor can be obtained analytically by integrating
\begin{equation}
\int_{a_i}^a\frac{d\tilde{a}}{\tilde{a} \sqrt{\frac{\Omega_r}{\tilde{a}^4} + \frac{\Omega_m}{\tilde{a}^3}}}=H_0 t,
\end{equation}
which, setting $a_i=0$, yields the exact solution
\begin{equation}
H_0 t = \frac{2}{3\Omega_m^2}
\left[
\left( a \Omega_m - 2\Omega_r \right)
\sqrt{a \Omega_m + \Omega_r}
+ 2\Omega_r^{3/2}
\right].
\label{eq:twofluid_solution}
\end{equation}
This expression provides the reference evolution for the standard radiation--matter cosmology.

Instead, in the presence of our new additional barotropic fluid, the evolution of the scale factor can only be determined numerically by solving the equation
\begin{equation}
\int_0^a \frac{d\tilde{a}}{\tilde{a} \sqrt{
\frac{\Omega_r}{\tilde{a}^4}
+ \frac{\Omega_m}{\tilde{a}^3}
+ \frac{\Omega_s}{\tilde{a}^{3(1+\omega_s)}
}}}=H_0 t.
\end{equation}

\begin{figure*}
\centering

\subfigure[]{%
    \includegraphics[width=0.52\textwidth,clip]{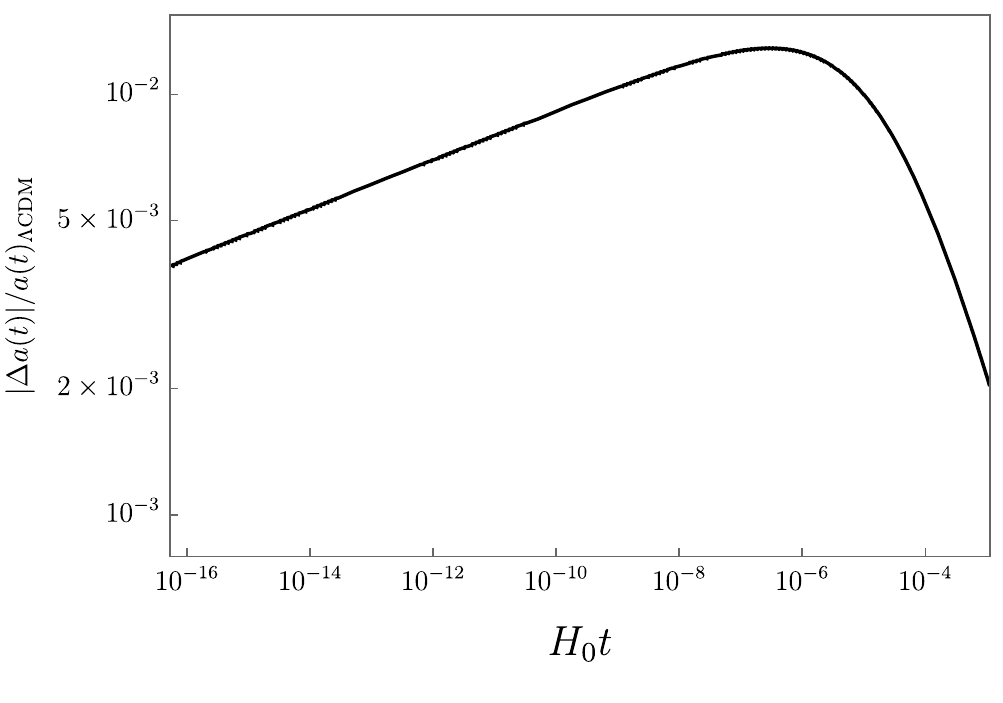}
    \label{fig:3vs2real}
}%
\subfigure[]{%
    \includegraphics[width=0.5\textwidth,clip]{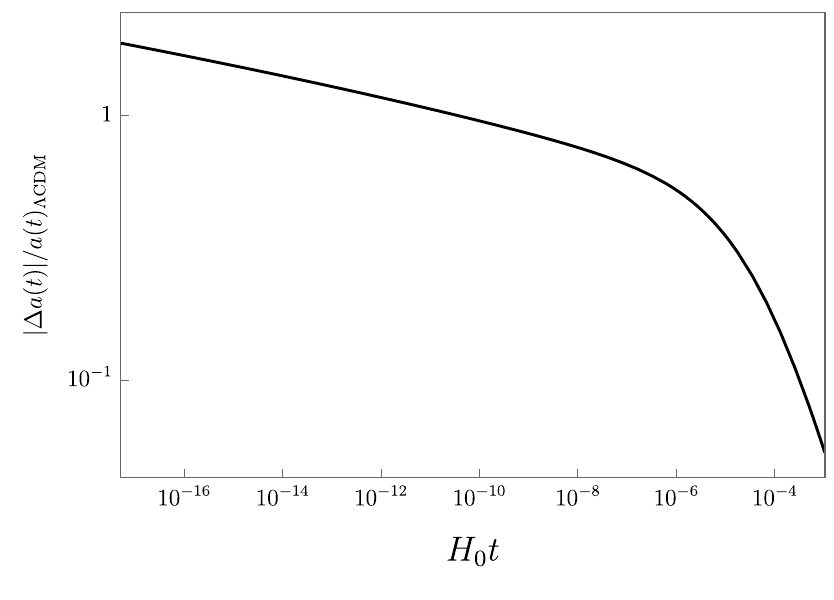}
    \label{fig:3vs2}
}

\caption{
Relative difference between the scale factor obtained from the numerical three-fluid solution and the analytical two-fluid $\Lambda$CDM solution. 
The left panel corresponds to the cosmological parameters derived in Ref.~\cite{Carloni:2025jlk}, with negligible deviations at early-times. 
The right panel refers to the case $\omega_s = 0.4$ and $\Omega_s = 8 \times 10^{-5}$, corresponding to the upper prior bounds of Ref.~\cite{Carloni:2025jlk}, where significant deviations arise at early-times due to the dominance of the additional component.
}
\label{fig:combined}
\end{figure*}

When $\omega_s < 1/3$, the dimensionless energy density of the additional component scales as
\begin{equation}
\Omega_s(a) \propto a^{-3(1+\omega_s)},
\end{equation}
which decreases more slowly than radiation. As a consequence, the radiation component dominates the expansion at sufficiently early-times.

This behavior is shown in the left panel of Fig.~\ref{fig:3vs2real}, where the numerical solution is nearly indistinguishable from the analytical two-fluid solution over the entire early-time range. 

This confirms that, in this regime, the additional component introduces only a correction that does not modify the radiation-dominated epoch, suggesting that the fluid remains consistent with current observational constraints.

For $\omega_s > 1/3$, the scaling of matter with pressure becomes steeper than that of radiation,
\begin{equation}
\Omega_s(a) \propto a^{-3(1+\omega_s)}, \qquad 3(1+\omega_s) > 4,
\end{equation}
and therefore the new fluid dominates the expansion at sufficiently early-times.

Here, the early-time behavior of the scale factor is provided by
\begin{equation}
a(t) \propto t^{\frac{2}{3(1+\omega_s)}},
\end{equation}
which differs from the standard radiation law $a(t) \propto t^{1/2}$.

As shown in the right panel of Fig.~\ref{fig:3vs2}, the numerical solution departs significantly from the analytical two-fluid result already at very early-times.

The comparison highlights a clear transition at $\omega_s = 1/3$. For $\omega_s < 1/3$, the additional component is subdominant in the asymptotic past and only affects the expansion at intermediate times. Conversely, for $\omega_s > 1/3$, it dominates the early Universe and replaces radiation as the leading contribution to the expansion rate.

This transition provides a direct diagnostic of the role of matter with pressure in shaping the early-time cosmological dynamics.

\section{Early-time consequences of our novel fluid}\label{early}

At early-times, cosmology in the presence of one more fluid yields more significant modifications of the Hubble parameter.  The corresponding Friedmann equations, in fact, may change, providing evidence toward an increase in the Hubble constant today, alleviating the cosmic tension. Moving from the $H_0$ tension, we here explore how Eq. \eqref{eq:rhos_standard} can change the sound speed and Big Bang Nucleosynthesis (BBN), showing that its influence, albeit quite relevant in fixing the tension, does not influence the growth of matter perturbations, implying no direct modifications of it, in analogy to radiation.

\subsection{The Hubble tension}

This scenario provides a novel early-time resolution of the Hubble tension, reducing it from $4.1\sigma$ to $2.4\sigma$. Within this framework, the standard two-fluid cosmology is generalized to a three-fluid description, in which the early Universe consists of photons, baryons, and matter with pressure.

The mechanism responsible for alleviating the Hubble tension relies on an enhanced early-time expansion rate $H(z)$, driven by the additional fluid component with $\omega_s$ and $\Omega_s$ representing the barotropic factor and the dimensionless density of the fluid, respectively. 

The new component leads to a reduction of the comoving sound horizon
\begin{equation}
r_s(z_\star) = \int_{z_\star}^{\infty} \frac{c_s(z)}{H(z)}dz,
\end{equation}
where $c_s(z)$ is the sound speed, and $z_\star$ indicates the redshift of recombination.

Moreover, since in our picture matter with pressure is always present, and contributes to the early-time expansion, it is coupled to the photon--baryon plasma, modifying the sound speed before the recombination as
\begin{equation}
c_s^2 = c^2\left[\frac{1 + 3\omega_s W}{3\left(1 + R + W\right)}\right].
\end{equation}
Here, $R \equiv 3\rho_b/(4\rho_\gamma)$ and $W \equiv 3(1+\omega_s)\rho_s/(4\rho_\gamma)$, with $\rho_b$, $\rho_s$, and $\rho_\gamma$ denoting the energy densities of baryons, matter with pressure, and photons, respectively.

\begin{figure}
\includegraphics[width=0.50\textwidth,clip]{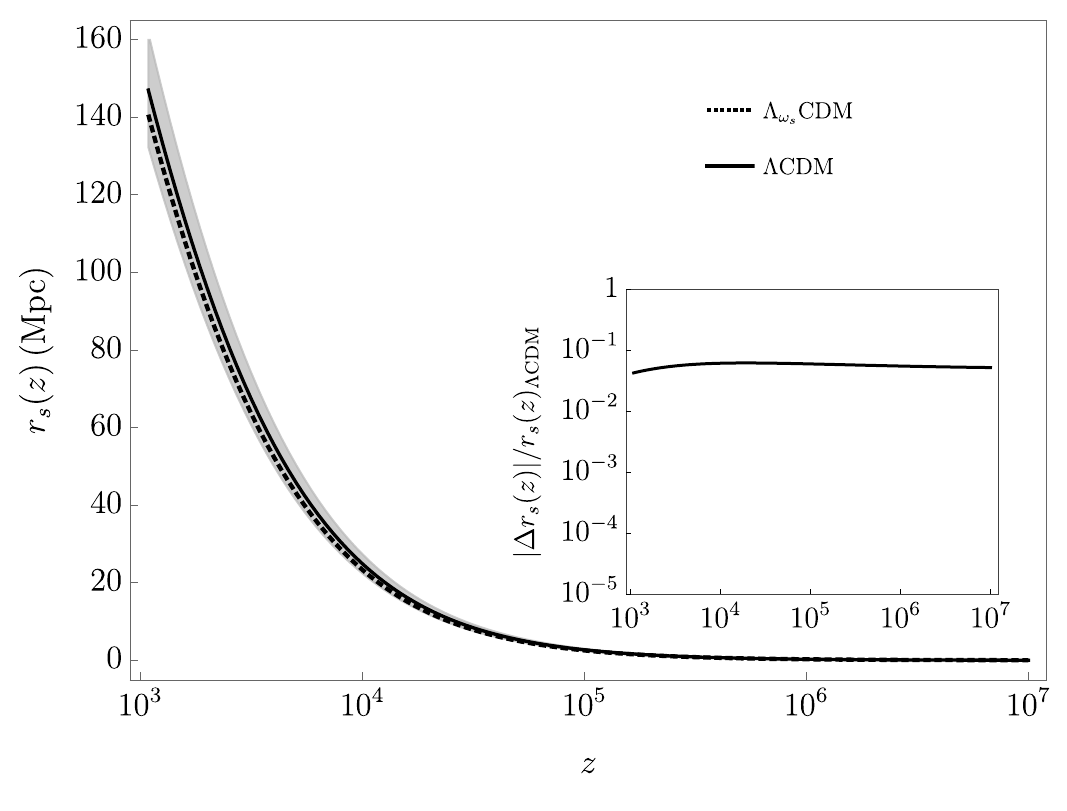}
\caption{Comoving sound horizon $r_s(z)$ for $\Lambda_{\omega_s}$CDM and $\Lambda$CDM. 
The shaded region indicates a $\pm 10\%$ band around the $\Lambda$CDM scenario. 
The inset displays the relative difference $|\Delta r_s(z)|/r_s(z)_{\Lambda\mathrm{CDM}}$. 
The cosmological parameter values are the same as those used in Fig.~\ref{fig:Omegatot}.}
\label{fig:rslcdm}
\end{figure}

In Ref.~\cite{Carloni:2025jlk}, we confirmed that the comoving sound horizon is reduced with respect to the $\Lambda$CDM model. As shown in Fig.~\ref{fig:rslcdm}, $r_s$ lies within the $10\%$ band of the $\Lambda$CDM prediction, but not within the $5\%$ band. The reduced value of $r_s$ implies a larger inferred $H_0$, thus alleviating the Hubble tension.

In the same analysis, we also verified that the inclusion of matter with pressure does not affect BBN, yielding $Y_p = 0.2478 \pm 0.0021$, in agreement with the value $Y_p = 0.2467 \pm 0.0002$ inferred from the Planck determination of the baryon density. This result is also consistent with the recent measurement of the primordial helium abundance \cite{Aver:2026dxv}, $Y_p = 0.2458 \pm 0.0013$, corresponding to a discrepancy of only $0.22 \sigma$ between the two determinations.

\subsection{Linear growth of matter perturbations}

The growth of cosmic structures provides a direct probe of both the expansion history and the clustering properties of the fluid components. In the concordance paradigm, cold dark matter perturbations grow via gravitational instability, giving rise to the observed LSS. The presence of additional components can modify this evolution either through their clustering properties or through their impact on the background dynamics.

In the $\Lambda_{\omega_s}$CDM model, we introduce an additional barotropic fluid characterized by a constant equation of state $P_s = \omega_s \rho_s$, with $\omega_s > 0$.

Here, we consider the results obtained in Ref.~\cite{Carloni:2025jlk} by combining CMB, DESI DR2, and Pantheon+\&SH0ES data, which yield $\omega_s = 0.294_{-0.004(0.023)}^{+0.014(0.015)}$ and $10^{5}\Omega_s = 1.62_{-0.56(0.91)}^{+0.36(1.02)}$.

In this scenario, we have demonstrated that matter with pressure contributes significantly to the total energy density only at early-times, while it does not behave as a clustering species. In particular, although perturbations in this fluid are present, they do not develop a growing mode capable of sourcing structure formation. This behavior is analogous to that of relativistic species.

Under the assumption that only matter clusters, the evolution of the linear matter density contrast, $\delta \equiv \delta_m / \rho_m$, is governed by the standard equation  
\begin{equation}
\ddot{\delta} + 2H \dot{\delta} - 4\pi G \rho_m \delta = 0.
\end{equation}
By adopting the scale factor $a$ as the independent variable, this equation can be rewritten as  
\begin{equation}
\delta'' + \left(\frac{3}{a} + \frac{H'}{H}\right)\delta'
- \frac{3\Omega_m}{2 a^5 E^2}\delta = 0,
\label{eq:deltaa}
\end{equation}
where a prime denotes differentiation with respect to $a$, and $E(a) \equiv H(a)/H_0$.

\begin{figure*}
\centering

\subfigure[]{%
    \includegraphics[width=0.5\textwidth,clip]{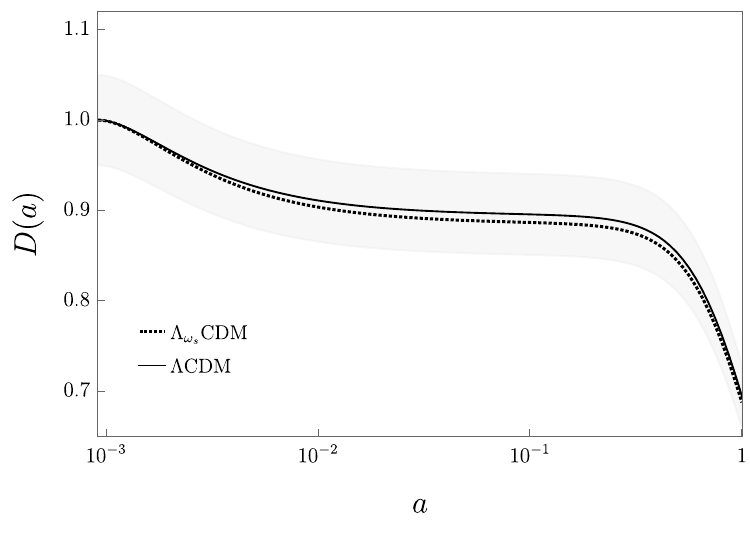}
    \label{fig:Da_left}
}%
\subfigure[]{%
    \includegraphics[width=0.5\textwidth,clip]{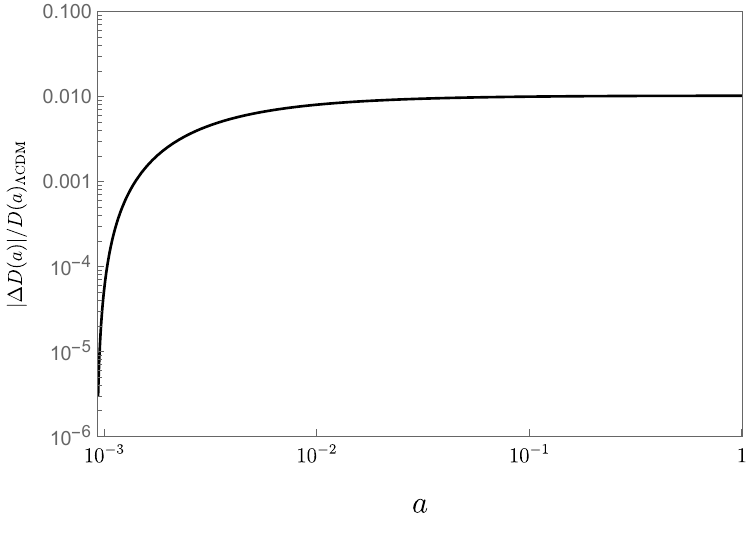}
    \label{fig:Da_right}
}

\caption{
Evolution of the growth function $D(a)$ in the $\Lambda_{\omega_s}$CDM model compared to the $\Lambda$CDM model. 
The left panel shows the evolution of $D(a)$, where the dashed curve corresponds to the $\Lambda_{\omega_s}$CDM model and the solid curve to the $\Lambda$CDM model, with the shaded region indicating a $\pm 5\%$ variation around the $\Lambda$CDM prediction. 
The right panel shows the relative difference $|\Delta D(a)|/D(a)_{\Lambda\mathrm{CDM}}$.
}
\label{fig:D(a)}
\end{figure*}

Introducing the growth variable $D(a) \equiv \delta/a$, the equation can be recast in the form  
\begin{align}
&D'' + D'\left[\frac{5}{a} + \frac{(\ln E^2)'}{2}\right]\nonumber \\
&+ \frac{D}{a} \left[
\frac{3}{a}\left(1 - \frac{\Omega_m}{2a^3 E^2}\right)
+ \frac{(\ln E^2)'}{2}
\right] = 0.
\end{align}
We solve this equation by imposing the boundary conditions $D(a_{\mathrm{LSS}})=1$ and $D'(a_{\mathrm{LSS}})=0$, where $a_{\mathrm{LSS}} = (1+z_{\mathrm{LSS}})^{-1}$ and $z_{\mathrm{LSS}} \simeq 1089$. The results, displayed in Fig.~\ref{fig:D(a)}, indicate that matter perturbations do not significantly deviate from the $\Lambda$CDM prediction within the considered cosmological scenario, exhibiting deviations below the $5\%$ level with respect to the concordance paradigm itself.

It is also convenient to describe structure growth through the logarithmic growth rate  
\begin{equation}
f(a) \equiv \frac{d \ln \delta}{d \ln a}.
\end{equation}
Combining this definition with the evolution equation for $\delta(a)$, we get the first-order differential equation  
\begin{equation}
f' + \frac{f^2}{a}
+ \left(\frac{2}{a} + \frac{E'}{E}\right) f
= \frac{3}{2}\frac{\Omega_{m}}{a^4 E^2}.
\end{equation}

\begin{figure*}
\centering

\subfigure[]{%
    \includegraphics[width=0.5\textwidth,clip]{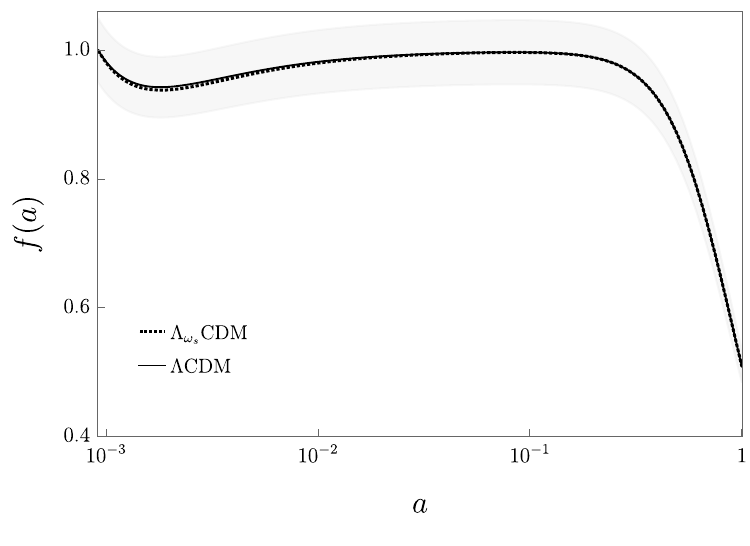}
    \label{fig:fa_left}
}%
\subfigure[]{%
    \includegraphics[width=0.5\textwidth,clip]{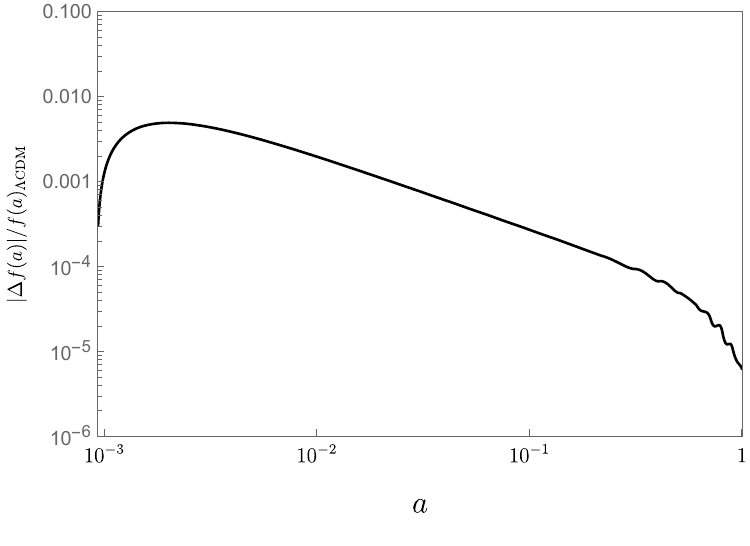}
    \label{fig:fa_right}
}

\caption{
Evolution of the growth rate $f(a)$ in the $\Lambda_{\omega_s}$CDM model compared to the $\Lambda$CDM model. 
The left panel shows the evolution of $f(a)$, where the dashed curve corresponds to the $\Lambda_{\omega_s}$CDM model and the solid curve to the $\Lambda$CDM model, with the shaded region indicating a $\pm 5\%$ variation around the $\Lambda$CDM prediction. 
The right panel shows the relative difference $|\Delta f(a)|/f(a)_{\Lambda\mathrm{CDM}}$.
}
\label{fig:f(a)}
\end{figure*}

As shown in Fig.~\ref{fig:f(a)}, the deviations from the $\Lambda$CDM prediction remain within the $\pm 5\%$ band throughout the entire evolution, indicating that the contribution of the matter with pressure component to structure growth is subdominant.

Finally, in order to assess the impact of our new fluid on a quantity directly related to LSS observations, we consider the observable $f\sigma_8(a)$, defined as
\begin{equation}
f\sigma_8(a)=f(a)\,\sigma_8(a),
\end{equation}
where the evolution of $\sigma_8$ is given by
\begin{equation}
\sigma_8(a)=\sigma_{8}\,\frac{\delta(a)}{\delta(1)}.
\end{equation}

In our analysis, we adopt $\sigma_{8}=0.829$ for the $\Lambda_{\omega_s}$CDM model and $\sigma_{8}=0.808$ for the $\Lambda$CDM case, while keeping $\Omega_{m}=0.296$ fixed \cite{Carloni:2025jlk}. 

\begin{figure*}
\centering

\subfigure[]{%
    \includegraphics[width=0.5\textwidth,clip]{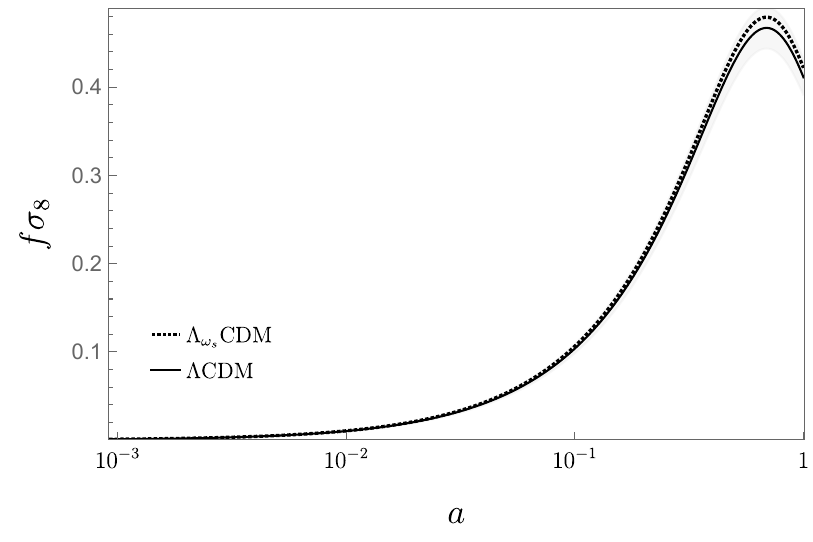}
    \label{fig:fs8_left}
}%
\subfigure[]{%
    \includegraphics[width=0.5\textwidth,clip]{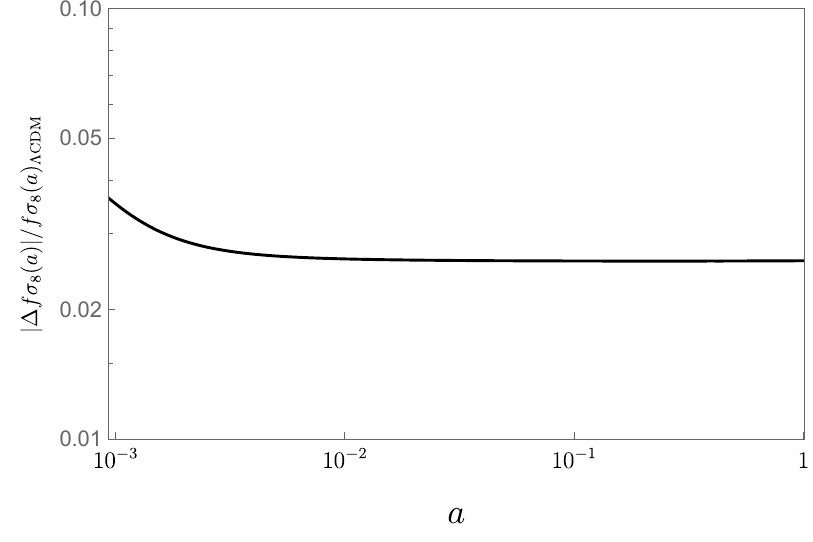}
    \label{fig:fs8_right}
}

\caption{
Evolution of $f\sigma_8(a)$ in the $\Lambda_{\omega_s}$CDM model compared to the $\Lambda$CDM model. 
The left panel shows the evolution of $f\sigma_8(a)$, where the dashed curve corresponds to the $\Lambda_{\omega_s}$CDM model and the solid curve to the $\Lambda$CDM model, with the shaded region indicating a $\pm 5\%$ variation around the $\Lambda$CDM prediction. 
The right panel shows the relative difference $|\Delta f\sigma_8(a)|/f\sigma{_8}(a)_{\Lambda\mathrm{CDM}}$.
}
\label{fig:fs8}
\end{figure*}

The results are depicted in Fig.~\ref{fig:fs8}. The upper panel indicates the evolution of $f\sigma_8(a)$ for the two cosmological models, while the lower panel presents the corresponding relative deviation. As clearly seen, the difference between the $\Lambda_{\omega_s}$CDM and $\Lambda$CDM predictions remains small over the entire range of the scale factor. In particular, the relative deviation stays at the percent level and is always contained within the $\pm 5\%$ gray band, despite the different values of $\sigma_{8}$ adopted in the two models.

This behavior confirms that the presence of the additional component does not induce any significant modification in the growth of matter perturbations. Indeed, although the new fluid affects the background expansion rate, especially at early-times, it does not cluster and therefore does not contribute directly to the source term governing the evolution of density perturbations. As a consequence, its impact on observables related to structure formation remains subdominant.

We therefore conclude that the growth of matter perturbations in the $\Lambda_{\omega_s}$CDM model closely follows the standard $\Lambda$CDM behavior. All considered quantities, including $D(a)$, $f(a)$, and $f\sigma_8(a)$, exhibit only small deviations that remain within observationally acceptable limits. This indicates that the introduction of the new non-clustering fluid preserves the predictions of the $\Lambda$CDM model at the level of linear structure formation.


\section{Bayesian analysis}\label{Sec.3}

We perform MCMCM analysis combining various probes to analyze the characteristics of $\Lambda_{\omega_s}$CDM cosmology. To this end, we extend the analysis of Ref.~\cite{Carloni:2025jlk}, originally aimed at alleviating the Hubble tension, in order to explore the behavior of the additional fluid under different dataset combinations. The model is implemented in a modified version of the \texttt{CLASS} Boltzmann code \footnote{\url{https://github.com/YouriCarloni/LwsCLASS.git}}, and parameter constraints are obtained using \texttt{MontePython}\footnote{\url{https://github.com/brinckmann/montepython_public.git}}.

We consider early- and late-time data, including the SH0ES prior $H_0=73.04\pm 1.04$ km/s/Mpc on SNe Ia. In addition, in line with Planck Collaboration analysis, we specified two massless neutrino species and one massive species with $m_\nu = 0.06\,\mathrm{eV}$, yielding to $N_{\rm eff}=3.046$ \cite{SimonsObservatory:2018koc}.

The datasets considered are described below.
\begin{itemize}
    \item[-]CMB \textit{Planck} 2018: We employ the \textit{Planck} 2018 CMB dataset, which comprises high-multipole TT, TE, and EE spectra, low-multipole TT and EE spectra, as well as lensing likelihoods. For the high-$\ell$ regime, we adopt the \texttt{Plik} likelihood \cite{Planck:2019nip}, incorporating TT measurements over the range $30 \le \ell \le 2508$ and TE/EE measurements within $30 \le \ell \le 1996$. At low multipoles, the TT likelihood in the interval $2 \le \ell \le 29$ is obtained via the \texttt{Commander} algorithm, whereas the EE likelihood over the same multipole range is derived using the \texttt{SimAll} likelihood \cite{Planck:2018vyg, Planck:2018yye}. Additionally, we include the CMB lensing likelihood reconstructed using the \texttt{Plik} pipeline, following the procedure outlined in Ref. \cite{Planck:2018lbu}.
    
    \item[-]DESI DR2: BAO measurements are given by DESI DR2 dataset reported in Tab.~\ref{tab:DESIDR2BAO}, which includes a total of $N_D = 13$ data points: six determinations of the transverse comoving distance $D_{\rm M}$, six measurements of the radial distance $D_{\rm H}$, and one estimate of the volume-averaged distance $D_{\rm V}$. All distances are normalized to the comoving sound horizon at the baryon drag epoch~\cite{DESI:2025zgx}. The corresponding BAO observables are defined as
\begin{subequations}
\begin{align}
\frac{D_{\rm M}(z)}{r_{\rm d}} &= \frac{c}{r_{\rm d}} \int_0^{z} \frac{dz'}{H(z')}, \\
\frac{D_{\rm H}(z)}{r_{\rm d}} &= \frac{c}{r_{\rm d},H(z)}, \\
\frac{D_{\rm V}(z)}{r_{\rm d}} &= \frac{\left[z,D_{\rm H}(z),D_{\rm M}^2(z)\right]^{1/3}}{r_{\rm d}},
\end{align}
\end{subequations}
where $r_{\rm d}$ denotes the comoving sound horizon at the drag epoch.

The BAO likelihood is then constructed as
\begin{equation}
\ln \mathcal{L}_{\rm BAO} \propto -\frac{1}{2} \sum_{i=1}^{N_D} \Delta Y_i^{\rm T} C^{-1} \Delta Y_i,
\end{equation}
with $\Delta Y_i = Y_i - Y(z_i)$. Here, $Y_i$ represents the DR2 observational data, while $Y(z_i) \in \left\{ D_{\rm M}(z_i)/r_{\rm d},, D_{\rm H}(z_i)/r_{\rm d},, D_{\rm V}(z_i)/r_{\rm d} \right\}$ denotes the corresponding theoretical predictions. The matrix $C$ is the associated covariance matrix\footnote{\url{https://github.com/LauraHerold/MontePython_desilike}}
.

\begin{table*}[htb!]
\centering
\setlength{\tabcolsep}{2.5em}
\renewcommand{\arraystretch}{1.1}
\begin{tabular}{|l|c|c|c|c|}
\hline\hline
Tracer     & $z_{\rm eff}$ & $D_{\rm M}/r_{\rm d}$ & $D_{\rm H}/r_{\rm d}$ & $D_{\rm V}/r_{\rm d}$ \\
\hline
\hline
BGS & $0.295$ & $-$ & $-$ & $7.942\pm 0.075$  \\
LRG1 & $0.510$ & $13.588\pm 0.167$ & $21.863\pm 0.425$ & $-$  \\
LRG2 & $0.706$ & $17.351\pm 0.177$ & $19.455\pm 0.330$ & $-$  \\
LRG3+ELG1 & $0.934$ & $21.576\pm 0.152$ & $17.641\pm 0.193$ & $-$ \\
ELG2 & $1.321$ & $27.601\pm 0.318$ & $14.176\pm 0.221$ & $-$  \\
QSO & $1.484$ & $30.512\pm 0.760$ & $12.817\pm 0.516$ & $-$  \\
Lya QSO & $2.330$ & $38.988\pm 0.531$ & $8.632\pm 0.101$ & $-$ \\
\hline
\hline
\end{tabular}
\caption{The DESI DR2 data points used in this analysis, for the bright galaxy survey (BGS), luminous red galaxies (LRG), emission line galaxies (ELG), quasars (QSO), Lyman-$\alpha$ forest quasars (Ly$\alpha$ QSO), and the combined LRG+ELG sample \cite{DESI:2025zgx}.}
\label{tab:DESIDR2BAO}
\end{table*}

\item[-]Pantheon: The Pantheon dataset employed in this analysis is described in Ref.~\cite{Pan-STARRS1:2017jku}. It consists of a compilation of 1048 SNe Ia, spanning the redshift interval $0.01 < z < 2.3$, obtained from the surveys included in the Pantheon sample after applying all selection criteria relevant for cosmological analyses. The catalog provides well-calibrated light curves, redshifts, and classifications, along with the mean redshift for each subsample.

Following Ref.~\cite{Pan-STARRS1:2017jku}, the likelihood is constructed from the vector of distance modulus residuals, defined as $\Delta \vec{\mu} = \vec{\mu}_{\rm obs} - \vec{\mu}_{\rm th}$. The observed distance modulus for each supernova is modeled as $\mu_{{\rm obs},i} = m_{B,i} - M$, where $m_{B,i}$ denotes the measured apparent magnitude of the $i$-th SNe Ia and $M$ is a nuisance parameter corresponding to the absolute magnitude. The theoretical prediction is given by
\begin{equation}
\mu_{\rm th}(z_i) = 5 \log_{10} D_{\rm L}(z_i) + 25,
\end{equation}
with $D_{\rm L}(z)$ denoting the luminosity distance expressed in units of Mpc, and the index $i$ labels each supernova in the sample. The likelihood can then be written as
\begin{equation}
\ln \mathcal{L}_{\rm SNe} = -\frac{1}{2} \Delta \vec{\mu}^{\,T} C^{-1} \Delta \vec{\mu},
\end{equation}
where $C$ represents the full covariance matrix, including both statistical and systematic uncertainties.
\item[-] OHD: We consider OHD data, consisting of $N_{\rm O}=32$ measurements of $H(z)$ spanning the redshift range $z \in [0.07,\,1.965]$, as reported in Tab.~\ref{tab:OHD}. Although these measurements are affected by relatively large uncertainties, they are obtained through a fully model-independent technique based on the \emph{cosmic chronometer} method~\cite{Jimenez:2001gg}. This approach exploits passively evolving galaxies, for which the Hubble parameter is estimated via the differential age method, by comparing the age difference between galaxy pairs formed at the same epoch but observed at slightly different redshifts, according to
\begin{equation}
H(z) = -\frac{1}{1+z}\,\frac{\Delta z}{\Delta t}.
\end{equation}
The model parameters are constrained by maximizing the corresponding likelihood 
\begin{equation}
\label{loglikeOHD}
\ln \mathcal{L}_{\rm OHD} \propto -\frac{1}{2} \sum_{i=1}^{N_{\rm O}} \left[ \frac{H_i - H(z_i)}{\sigma_{H_i}} \right]^2,
\end{equation}
where $\sigma_{H_i}$ denotes the uncertainty associated with each measurement of $H(z)$.

\begin{table}[htb!]
\centering
\setlength{\tabcolsep}{1.5em}
\renewcommand{\arraystretch}{1.1}
\begin{tabular}{|c|c|c|}
   \hline\hline
    $z$     &$H(z)$ &  References \\
            &[km/s/Mpc]&\\
    \hline\hline
    0.07  & $69.0\pm 19.6$ & \cite{Zhang:2012mp} \\
    0.09    & $69.0 \pm12.0$  & \cite{Jimenez:2001gg} \\
    0.12    & $68.6\pm26.2$  & \cite{Zhang:2012mp} \\
    0.17    & $83.0\pm8.0$   & \cite{Simon:2004tf} \\
    0.179   & $75.0  \pm 4.0$   & \cite{Moresco:2012jh} \\
    0.199   & $75.0\pm5.0$   & \cite{Moresco:2012jh} \\
    0.20    & $72.9\pm29.6$  & \cite{Zhang:2012mp} \\
    0.27    & $77.0\pm14.0$  & \cite{Simon:2004tf} \\
    0.28    & $88.8\pm36.6$  & \cite{Zhang:2012mp} \\
    0.352  & $83.0\pm14.0$  & \cite{Moresco:2016mzx} \\
    0.38  & $83.0\pm13.5$  & \cite{Moresco:2016mzx} \\
    0.4     & $95.0\pm17.0$  & \cite{Simon:2004tf} \\
    0.4004  & $77.0\pm10.2$  & \cite{Moresco:2016mzx} \\
    0.425  & $87.1\pm11.2$  & \cite{Moresco:2016mzx} \\
    0.445  & $92.8 \pm12.9$  & \cite{Moresco:2016mzx} \\
    0.47    & $89.0\pm23.0$     & \cite{Ratsimbazafy:2017vga}\\
    0.4783  & $80.9\pm9.0$   & \cite{Moresco:2016mzx} \\
    0.48    & $97.0\pm62.0$  & \cite{Stern:2009ep} \\
    0.593   & $104.0\pm13.0$  & \cite{Moresco:2012jh} \\
    0.68    & $92.0\pm8.0$   & \cite{Moresco:2012jh} \\
    0.75    & $98.8\pm33.6$     & \cite{Borghi:2021rft}\\
    0.781  & $105.0\pm12.0$  & \cite{Moresco:2012jh} \\
    0.875   & $125.0\pm17.0$  & \cite{Moresco:2012jh} \\
    0.88    & $90.0\pm40.0$  & \cite{Stern:2009ep} \\
    0.9     & $117.0\pm23.0$  & \cite{Simon:2004tf} \\
    1.037   & $154.0\pm20.0$  & \cite{Moresco:2012jh} \\
    1.3     & $168.0\pm17.0$  & \cite{Simon:2004tf} \\
    1.363   & $160.0 \pm 33.6$  & \cite{Moresco:2015cya} \\
    1.43    & $177.0\pm18.0$  & \cite{Simon:2004tf} \\
    1.53    & $140.0\pm14.0$  & \cite{Simon:2004tf} \\
    1.75    & $202.0\pm40.0$  & \cite{Simon:2004tf} \\
    1.965   & $186.5 \pm 50.4$  & \cite{Moresco:2015cya} \\
\hline\hline
\end{tabular}
\caption{Compilation of OHD measurements, including redshift values (first column), corresponding $H(z)$ estimates with uncertainties (second column), and associated references (third column)}
\label{tab:OHD}
\end{table}
\end{itemize}

\begin{table*}
\begin{tabular*}{\linewidth}{@{\extracolsep{\fill}}lcccc}
\hline\hline
Parameter & CMB & CMB+BAO & CMB+P & CMB+BAO+P \\
\hline\hline

$100\,\omega_b$ 
& $2.237^{+0.015(0.031)}_{-0.015(0.031)}$ 
& $2.252^{+0.014(0.027)}_{-0.013(0.028)}$
& $2.239^{+0.015(0.029)}_{-0.015(0.030)}$
& $2.252^{+0.014(0.027)}_{-0.013(0.028)}$ \\

$\omega_{\rm cdm}$ 
& $0.1201^{+0.0012(0.0025)}_{-0.0013(0.0025)}$
& $0.1182^{+0.0007(0.0023)}_{-0.0012(0.0019)}$
& $0.1200^{+0.0012(0.0027)}_{-0.0013(0.0026)}$
& $0.1180^{+0.0007(0.0020)}_{-0.0010(0.0017)}$ \\

$H_0$~(km/s/Mpc) 
& $67.53^{+0.56(1.19)}_{-0.61(1.17)}$
& $68.58^{+0.32(0.80)}_{-0.40(0.75)}$
& $67.64^{+0.55(1.22)}_{-0.61(1.16)}$
& $68.52^{+0.32(0.66)}_{-0.34(0.66)}$ \\

$10^{9}A_s$ 
& $2.106^{+0.031(0.065)}_{-0.0330.064)}$
& $2.124^{+0.031(0.067)}_{-0.034(0.064)}$
& $2.106^{+0.030(0.064)}_{-0.033(0.064)}$
& $2.124^{+0.034(0.066)}_{-0.033(0.066)}$ \\

$n_s$ 
& $0.9665^{+0.0047(0.0091)}_{-0.0044(0.0091)}$
& $0.9726^{+0.0038(0.0083)}_{-0.0043(0.0082)}$
& $0.9670^{+0.0044(0.0094)}_{-0.0048(0.0088)}$
& $0.9720^{+0.0037(0.0076)}_{-0.0038(0.0076)}$ \\

$\tau_{\rm reio}$ 
& $0.0549^{+0.0075(0.016)}_{-0.0081(0.016)}$
& $0.0608^{+0.0070(0.016)}_{-0.0081(0.015)}$
& $0.0549^{+0.0079(0.016)}_{-0.0078(0.016)}$
& $0.0610^{+0.0075(0.016)}_{-0.0079(0.016)}$ \\

$\omega_s$ 
& $< 0.22 \; (2\sigma)$
& N.C.
& $< 0.128 \; (1\sigma)$
& $<0.25 \; (2\sigma)$ \\

$10^{5}\Omega_{s}$  
& $< 9.25 \; (2\sigma)$
& N.C.
& $< 9.13 \; (2\sigma)$
& N.C. \\

\hline

$r_s(z_{\star})$~(Mpc) 
& $144.4^{+0.3(0.7)}_{-0.3(0.7)}$
& $144.7^{+0.5(0.7)}_{-0.2(1.0)}$
& $144.4^{+0.4(0.7)}_{-0.3(0.9)}$
& $144.8^{+0.4(0.6)}_{-0.2(0.7)}$ \\

$\sigma_8$ 
& $0.811^{+0.006(0.013)}_{-0.007(0.013)}$
& $0.810^{+0.007(0.013)}_{-0.007(0.013)}$
& $0.811^{+0.006(0.013)}_{-0.006(0.013)}$
& $0.809^{+0.007(0.013)}_{-0.007(0.013)}$ \\

$\Omega_m$
& $0.314^{+0.007(0.015)}_{-0.008(0.016)}$
& $0.300^{+0.004(0.008)}_{-0.004(0.008)}$
& $0.313^{+0.007(0.015)}_{-0.007(0.015)}$
& $0.301^{+0.004(0.008)}_{-0.004(0.008)}$ \\

$S_8$
& $0.830^{+0.013(0.025)}_{-0.013(0.025)}$

& $0.810^{+0.009(0.017)}_{-0.009(0.017)}$

& $0.828^{+0.012(0.025)}_{-0.012(0.024)}$

& $0.810^{+0.008(0.018)}_{-0.009(0.017)}$ \\

$100\,\theta_s$ 
& $1.0419^{+0.0003(0.0007)}_{-0.0003(0.0006)}$
& $1.0422^{+0.0003(0.0007)}_{-0.0003(0.0006)}$
& $1.0420^{+0.0003(0.0007)}_{-0.0003(0.0007)}$
& $1.0422^{+0.0003(0.0006)}_{-0.0003(0.0006)}$ \\

\hline\hline

\end{tabular*}

\caption{MCMC constraints on the $\Lambda_{\omega_s}$CDM model. Mean values with $1\sigma$ ($2\sigma$) uncertainties derived from CMB, CMB+BAO, CMB+Pantheon and CMB+BAO+Pantheon datasets.}
\label{tab:best}
\end{table*}

\begin{table}[t]
\centering
\resizebox{\columnwidth}{!}{%
\begin{tabular}{lcc}
\hline\hline
Parameter & CMB+BAO+P+SH0ES & CMB+P+SH0ES \\
\hline\hline

$100\,\omega_b$ 
& $2.268^{+0.014(0.028)}_{-0.014(0.028)}$
& $2.278^{+0.018(0.036)}_{-0.018(0.035)}$ \\

$\omega_{\rm cdm}$ 
& $0.1257^{+0.0027(0.0054)}_{-0.0028(0.0054)}$
& $0.1261^{+0.0030(0.0059)}_{-0.0030(0.0059)}$ \\

$H_0$~(km/s/Mpc) 
& $71.01^{+0.75(1.46)}_{-0.75(1.46)}$
& $71.46^{+1.07(1.95)}_{-0.94(2.02)}$ \\

$10^{9}A_s$ 
& $2.123^{+0.031(0.069)}_{-0.037(0.066)}$
& $2.121^{+0.033(0.072)}_{-0.038(0.070)}$ \\

$n_s$ 
& $0.9806^{+0.0044(0.0091)}_{-0.0047(0.0092)}$
& $0.9792^{+0.0052(0.0114)}_{-0.0060(0.0110)}$ \\

$\tau_{\rm reio}$ 
& $0.0599^{+0.0071(0.016)}_{-0.0083(0.015)}$
& $0.0615^{+0.0077(0.016)}_{-0.0085(0.016)}$ \\

$\omega_s$ 
& $0.290^{+0.017(0.021)}_{-0.007(0.028)}$
& $0.302^{+0.024(0.034)}_{-0.013(0.038)}$ \\

$10^{5}\Omega_{s}$  
& $1.47^{+0.35(1.14)}_{-0.62(0.94)}$
& $1.21^{+0.31(1.10)}_{-0.65(0.86)}$ \\

\hline

$r_s(z_{\star})$~(Mpc) 
& $140.9^{+1.3(2.5)}_{-1.3(2.5)}$
& $140.5^{+1.3(2.8)}_{-1.5(2.7)}$ \\

$\sigma_8$ 
& $0.823^{+0.008(0.016)}_{-0.008(0.016)}$
& $0.824^{+0.008(0.017)}_{-0.009(0.016)}$ \\

$\Omega_m$
& $0.296^{+0.004(0.007)}_{-0.004(0.007)}$
& $0.293^{+0.006(0.013)}_{-0.007(0.012)}$ \\

$S_8$
& $0.819^{+0.009(0.018)}_{-0.009(0.018)}$ 
& $0.814^{+0.012(0.024)}_{-0.012(0.023)}$ \\

$100\,\theta_s$ 
& $1.0431^{+0.0004(0.0008)}_{-0.0004(0.0008)}$
& $1.0430^{+0.0004(0.0009)}_{-0.0005(0.0009)}$ \\

\hline\hline
\end{tabular}%
}
\caption{MCMC constraints on the $\Lambda_{\omega_s}$CDM model. Mean values with $1\sigma$ ($2\sigma$) uncertainties derived including SH0ES prior, with and without BAO data.}
\label{tab:best_shoes}
\end{table}

\begin{table}
\centering
\scriptsize
\setlength{\tabcolsep}{3pt}
\renewcommand{\arraystretch}{1.1}
\begin{tabular}{lcc}
\hline\hline
Parameter & without GP & with GP \\
\hline\hline

\multicolumn{3}{c}{\textbf{without OHD}} \\

$\Omega_m$ 
& $0.298^{+0.008(0.017)}_{-0.009(0.017)}$
& $0.299^{+0.008(0.017)}_{-0.009(0.017)}$ \\

$H_0$ 
& $72.74^{+1.15(2.22)}_{-1.08(2.29)}$
& $72.73^{+1.15(2.24)}_{-1.09(2.28)}$ \\

$\omega_s$ 
& $0.241^{+0.027(0.083)}_{-0.039(0.061)}$
& $0.284^{+0.025(0.047)}_{-0.018(0.045)}$ \\

$10^{5}\Omega_s$  
& N.C.
& $1.49^{+0.47(0.93)}_{-0.50(0.95)}$ \\

$\chi^2_{\rm min}$  
& $1036$ & $1036$ \\

\hline

\multicolumn{3}{c}{\textbf{with OHD}} \\

$\Omega_m$ 
& $0.296^{+0.008(0.017)}_{-0.008(0.016)}$
& $0.296^{+0.008(0.017)}_{-0.009(0.016)}$ \\

$H_0$ 
& $71.61^{+1.00(1.91)}_{-0.95(1.94)}$
& $71.59^{+1.03(1.91)}_{-0.94(2.01)}$ \\

$\omega_s$ 
& $0.224^{+0.036(0.088)}_{-0.032(0.077)}$
& $0.266^{+0.037(0.069)}_{-0.015(0.098)}$ \\

$10^{5}\Omega_s$  
& N.C.
& $1.48^{+0.47(0.94)}_{-0.50(0.95)}$ \\

$\chi^2_{\rm min}$  
& $1055$ & $1055$ \\

\hline\hline
\end{tabular}
\caption{Constraints on the $\Lambda_{\omega_s}$CDM model from background data. Mean values with $1\sigma$ ($2\sigma$) uncertainties are obtained from Pantheon, BAO, and SH0ES data, with and without OHD, and with and without a Gaussian prior on $\Omega_s$. The OHD data contribute $\Delta\chi^2_{\rm min} \simeq 19$.}
\label{tab:background_full}
\end{table}

The contour plots obtained from our analyses are presented in Appendix~\ref{AppA}, while the mean parameter values and their corresponding uncertainties are reported in Tabs.~\ref{tab:best}, \ref{tab:best_shoes}, and \ref{tab:background_full}. The priors adopted are given in Tab. \ref{tab:priors}.

We assess chain convergence for the $\Lambda_{\omega_s}$CDM cosmology by requiring that the Gelman--Rubin statistic satisfies $R - 1 < 0.02$ \cite{Gelman1992}, following the same criterion adopted in our previous analysis \cite{Carloni:2025jlk}.

Our findings show that the additional fluid emerges only when this prior is included, suggesting that our cosmology prefers a higher value of $H_0$ compared to the $\Lambda$CDM model.

Specifically, in Tab.~\ref{tab:best}, we consider four combinations of datasets: CMB, CMB+BAO, CMB+Pantheon, and CMB+BAO+Pantheon. In all cases, the additional parameters $\omega_s$ and $\Omega_{s}$ remain unconstrained and tend toward the lower bounds of their prior distributions. This indicates that, since the fluid component vanishes in the limit $\Omega_{s} \to 0$, the $\Lambda_{\omega_s}$CDM cosmology is effectively excluded and the $\Lambda$CDM paradigm is recovered.

The same behavior is also observed in the EDE scenario. In particular, in the absence of SH0ES prior, both the cosmological models, which predict a higher value of $H_0$, are ruled out from the MCMC analysis.

Instead, when the SH0ES prior is included, Tab.~ \ref{tab:best_shoes}, the $\Lambda_{\omega_s}$CDM model emerges, with and without considering the BAO data. The values of the addition parameters are $\omega_s = 0.290^{+0.017(0.021)}_{-0.007(0.028)}$, $10^5\Omega_{s} = 1.47^{+0.35(1.14)}_{-0.62(0.94)}$, and $\omega_s = 0.302^{+0.024(0.034)}_{-0.013(0.038)}$, $10^5\Omega_{s} = 1.21^{+0.31(1.10)}_{-0.65(0.86)}$ in the two cases, respectively.

In Tab.~\ref{tab:chi2}, we report the individual contributions to the total $\chi^2$, from each cosmic probe in the combined analysis, suggesting their consistency within the global fit.

Finally, we consider only late-time probes, excluding CMB data, as shown in Tab.~\ref{tab:background_full}. 

We perform analyses with and without OHD data, and in each case, we consider both the inclusion and exclusion of a Gaussian prior on $\Omega_{s}$ to assess the impact on the results.

The Gaussian prior on $\Omega_{s}$ is given by $10^5\Omega_{s} = 1.62\pm 0.46$, and it is derived from the analysis presented in Ref.~\cite{Carloni:2025jlk}, where CMB, DESI DR2, and Pantheon+ \& SH0ES datasets are combined.

We observe that, without the Gaussian prior on $\Omega_{s}$, only $\omega_s$ is constrained, with $\omega_s = 0.224^{+0.036(0.088)}_{-0.032(0.077)}$ for analyses including OHD and $\omega_s = 0.241^{+0.027(0.083)}_{-0.039(0.061)}$ when OHD data are excluded. Instead, when we add the Gaussian prior on $\Omega_{s}$, both parameters are consistent with those obtained from the full dataset combination, yielding $\omega_s = 0.279^{+0.023(0.045)}_{-0.017(0.043)}$ and $10^5 \Omega_{s} = 1.46^{+0.45(0.90)}_{-0.49(0.92)}$ with OHD included, and $\omega_s =0.284^{+0.025(0.047)}_{-0.018(0.045)}$ and $10^5 \Omega_{s} = 1.49^{+0.47(0.93)}_{-0.50(0.95)}$ without OHD.

The results show no differences in $\chi^2_{\rm min}$ if we add the Gaussian prior on $\Omega_s$, indicating that the model is not excluded from a statistical point of view.

\begin{table}[htb!]
\centering
\begin{tabular}{lc}
\hline\hline
Parameter & Prior \\
\hline

$100\,\omega_b$ & $\mathcal{U}(1.8, 3)$ \\
$\omega_{\rm cdm}$ & $\mathcal{U}(0.1, 0.2)$ \\
$H_0$~(km/s/Mpc) & $\mathcal{U}(50, 80)$ \\
$10^{9}A_s$ & $\mathcal{U}(1.8, 3)$ \\
$n_s$ & $\mathcal{U}(0.9, 1.1)$ \\
$\tau_{\rm reio}$ & $\mathcal{U}(0.004, 0.12)$ \\
$\omega_s$ & $\mathcal{U}(0.0001, 5)$ \\
$10^{5}\Omega_s$ & $\mathcal{U}(0.0001, 10)$/$\mathcal{N}(0.0001, 10)$ \\

\hline\hline
\end{tabular}

\caption{Uniform priors on the cosmological parameters of the $\Lambda_{\omega_s}$CDM model. For the background probes only, we also consider the effect of a Gaussian prior on $\Omega_s$.}
\label{tab:priors}
\end{table}

The behavior of the $\Lambda_{\omega_s}$CDM model is physically well motivated.
When only early-time probes are included, the extra fluid remains unconstrained, and the parameters $\omega_s$ and $\Omega_s$ are pushed toward the edges of their priors. This is expected, as matter with pressure is assumed to be subdominant, and early-time data alone do not require its presence. In this limit, the model naturally recovers $\Lambda$CDM since $\Omega_s \to 0$.

Once the SH0ES prior is added, the degeneracy with $H_0$ is broken, and the fluid parameters emerge, giving a higher value of the Hubble constant. This is analogous, at the phenomenological level, to what happens in EDE models.

The difference with respect to EDE is deduced from the physical characteristics of our novel fluid. 

In EDE, the extra component is typically negligible until near recombination, so larger values of $H_0$ can be accommodated without strongly affecting the pre-recombination cosmology more than a frozen scalar field. 

In contrast, matter with pressure is always present and contributes non-negligibly already at early-times. 

Although subdominant, it modifies the total energy budget and can therefore produce a stronger impact on the CMB acoustic pattern. As a consequence, in the absence of an external $H_0$ prior, the likelihood does not naturally favor high values of $H_0$, and the fit remains close to the $\Lambda$CDM solution. This characteristic should not be regarded as a shortcoming of the model, but rather as a direct consequence of its physical properties.

Moreover, when only CMB data are considered, neither $\omega_s$ nor $\Omega_s$ is constrained. This is due to the fact that CMB measurements are sensitive not only to the background expansion, but also to the evolution of perturbations. In particular, matter with pressure becomes significant as the scale factor decreases, dominating over the dust at early-times, and this induces modifications to both the background and perturbation dynamics, affecting the CMB anisotropy spectrum.

Therefore, the $\Lambda_{\omega_s}$CDM model is more easily accommodated by late-time data, which can constrain at least the barotropic parameter $\omega_s$, while $\Omega_s$ remains not determined unless a Gaussian prior is imposed.

\begin{table}[htb!]
\centering
\begin{tabular}{lc}
\hline\hline
Dataset & $\chi^2_{\min}$ \\
\hline

Planck high-$\ell$ TT,TE,EE + low-$\ell$ TT,EE + lensing & $2778$ \\
BAO (DESI DR2) & $16$ \\
Pantheon & $1026$ \\
SH0ES prior & $4$ \\

\hline
Total & $3824$ \\
\hline\hline
\end{tabular}

\caption{Minimum $\chi^2$ contributions from each dataset in the $\Lambda_{\omega_s}$CDM model for the full combined analysis.}
\label{tab:chi2}
\end{table}

\section{Comparison between $\Lambda_{\omega_s}$CDM and Early Dark Energy}\label{sec6}

\begin{figure*}[htb!]
\includegraphics[width=\hsize,clip]{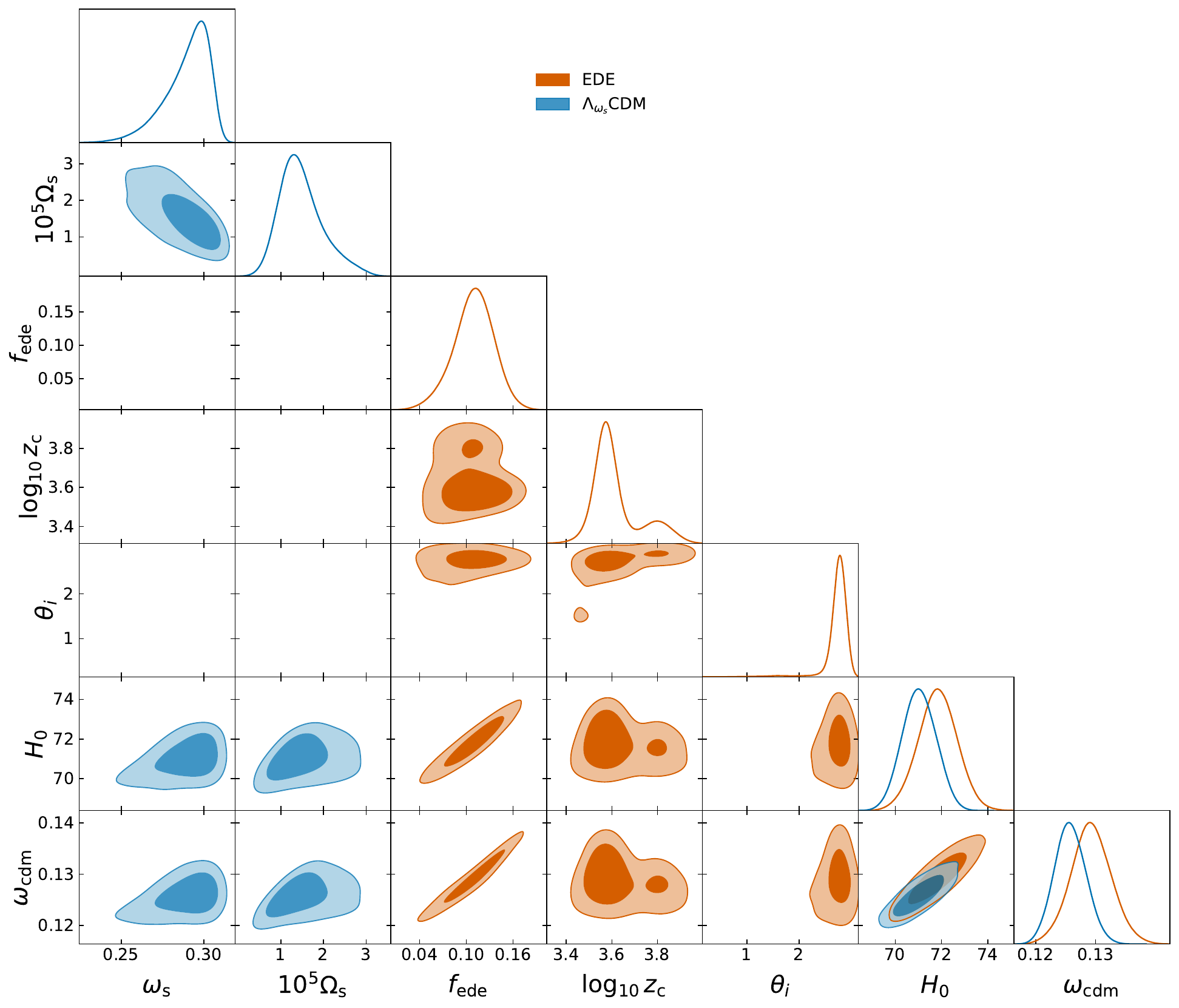}
\caption{Comparison between the EDE and $\Lambda_{\omega_s}$CDM models. The posterior distributions are obtained using \textit{Planck} 2018 CMB + BAO (DESI DR2) + Pantheon + SH0ES.}
\label{fig:edevslw}
\end{figure*}

Both the EDE and $\Lambda_{\omega_s}$CDM models can alleviate the $H_0$ tension, although they exhibit distinct physical characteristics. 

\begin{table}[htb!]
\centering
\resizebox{\columnwidth}{!}{%
\begin{tabular}{lcc}
\hline\hline
Parameter & $\Lambda_{\omega_s}$CDM & EDE \\
\hline\hline

$100\,\omega_b$ 
& $2.268^{+0.014(0.028)}_{-0.014(0.028)}$
& $2.284^{+0.020(0.044)}_{-0.022(0.043)}$ \\

$\omega_{\rm cdm}$ 
& $0.1257^{+0.0027(0.0054)}_{-0.0028(0.0054)}$
& $0.1294^{+0.0033(0.0068)}_{-0.0032(0.0065)}$ \\

$H_0$~(km/s/Mpc) 
& $71.01^{+0.75(1.46)}_{-0.75(1.46)}$
& $71.81^{+0.81(1.71)}_{-0.80(1.61)}$ \\

$10^{9}A_s$ 
& $2.123^{+0.031(0.069)}_{-0.037(0.066)}$
& $2.156^{+0.031(0.069)}_{-0.036(0.068)}$ \\

$n_s$ 
& $0.9806^{+0.0044(0.0091)}_{-0.0047(0.0092)}$
& $0.9898^{+0.0060(0.0126)}_{-0.0065(0.0126)}$ \\

$\tau_{\rm reio}$ 
& $0.0599^{+0.0071(0.016)}_{-0.0083(0.015)}$
& $0.0601^{+0.0072(0.0160)}_{-0.0084(0.0157)}$ \\

$\omega_s$ 
& $0.290^{+0.017(0.021)}_{-0.007(0.028)}$
& -- \\

$10^{5}\Omega_{s}$  
& $1.47^{+0.35(1.14)}_{-0.62(0.94)}$
& -- \\

$f_{\rm ede}$
& --
& $0.1098^{+0.0269(0.0506)}_{-0.0221(0.0528)}$ \\

$\theta_i$
& --
& $2.738^{+0.160(0.301)}_{-0.105(0.300)}$ \\

$\log_{10}(z_c)$
& --
& $3.609^{+0.028(\text{--})}_{-0.100(\text{--})}$ \\

\hline

$r_s(z_{\star})$~(Mpc) 
& $140.9^{+1.3(2.5)}_{-1.3(2.5)}$
& $139.0^{+1.4(3.1)}_{-1.5(3.0)}$ \\

$\sigma_8$ 
& $0.823^{+0.008(0.016)}_{-0.008(0.016)}$
& $0.836^{+0.0095(0.0192)}_{-0.0100(0.0193)}$ \\

$\Omega_m$
& $0.296^{+0.004(0.007)}_{-0.004(0.007)}$
& $0.2963^{+0.0033(0.0070)}_{-0.0036(0.0070)}$ \\

$S_8$
& $0.819^{+0.009(0.018)}_{-0.009(0.018)}$ 
& $0.831^{+0.011(0.022)}_{-0.011(0.022)}$ \\

$100\,\theta_s$ 
& $1.0431^{+0.0004(0.0008)}_{-0.0004(0.0008)}$
& $1.0415^{+0.00037(0.00073)}_{-0.00038(0.00074)}$ \\

\hline

$\chi^{2}_{\rm min}$ $(\Delta\chi^2_{\rm min}$) & $3824(8)$ & $3816(0)$  \\

AIC $(\Delta \mathrm{AIC})$ & $3840(6)$ & $3834(0)$\\

DIC $(\Delta \mathrm{DIC})$ & $3872(3)$ & $3869(0)$ \\

\hline\hline
\end{tabular}%
}
\caption{MCMC constraints on the $\Lambda_{\omega_s}$CDM and EDE models. Mean values with $1\sigma$ ($2\sigma$) uncertainties are obtained by combining \textit{Planck} 2018 CMB, BAO (DESI DR2), and Pantheon data, including SH0ES prior on $H_0$.}
\label{tab:comparison}
\end{table}

In particular, EDE is represented by an axion-like scalar field $\phi$ with a potential of the form
\begin{equation}
V(\phi)=V_0\left(1 - \cos(\phi/f)\right)^n,
\end{equation}
where $V_0 = m^2 f^2$, with $m$ denoting the axion mass and $f$ the decay constant \cite{Poulin:2018dzj}.

At early-times, the field is effectively frozen due to Hubble friction and behaves as a cosmological constant. As the expansion rate decreases and the Hubble parameter falls below a critical value, corresponding to a redshift $z_c = a_c^{-1} - 1$, the field begins to oscillate around the minimum of the potential. In this regime, it can be described as an effective fluid with equation of state $\omega_n = (n-1)/(n+1)$.

The model introduces three additional parameters: the mass of the field $m$, the decay constant $f$, and the initial field value $\theta_i = \phi_i/f$. From these quantities, we can derive the fractional energy density at the critical redshift, $f_{\rm ede}(z_c) \equiv \Omega_\phi(z_c)/\Omega(z_c)$, as well as the redshift $z_c$ itself. In Fig.~\ref{fig:edevslw}, we show the marginalized posterior distributions of $f_{\rm EDE}(z_c)$, $\log_{10}(z_c)$, and $\theta_i$, which fully characterize the additional parameter space of the model.

Moreover, for the potential scaling, we fix $n=3$, as this value is favored by observational data: it ensures a sufficiently rapid decay of EDE and provides a more effective alleviation of the $H_0$ tension \cite{Poulin:2018cxd}.

At this stage, we perform a statistical comparison between $\Lambda_{\omega_s}$CDM and EDE to establish the competitiveness of our novel model. 

As already shown in Ref.~\cite{Carloni:2025jlk}, $\Lambda_{\omega_s}$CDM is favored over $\Lambda$CDM by combining \textit{Planck} 2018 CMB, DESI DR2, and Pantheon+ \& SH0ES data. 

We now apply statistical tools to determine whether it can be considered comparable to EDE.

Specifically, we employ two widely used information criteria, {\it i.e.}, the Akaike Information Criterion (AIC) and the Deviance Information Criterion (DIC) \cite{Kunz:2006mc,Liddle:2007fy,Biesiada:2007um,Szydlowski:2005kv,Szydlowski:2006pz}, to assess the relative performance of our scenario.

These estimators, commonly adopted in statistical inference and model selection, are defined as
\begin{subequations}
\label{eq:AIC_DIC}
\begin{align}
& \mathrm{AIC} = -2\ln \mathcal{L}_{\max} + 2k, \\
& \mathrm{DIC} = -2\ln \mathcal{L}_{\max} + 2p,
\end{align}
\end{subequations}
where $\ln \mathcal{L}_{\max}$ indicates the maximum log-likelihood, $k$ is the number of free parameters, and $p = \langle -2\ln \mathcal{L} \rangle + 2\ln \mathcal{L}_{\max}$ represents the Bayesian complexity, with angular brackets denoting an average over the posterior distribution. The quantity $p$ can be interpreted as the effective number of parameters constrained by the data \cite{Liddle:2007fy}. 

Although AIC and DIC have a similar structure, they differ in how they account for model complexity.

From a statistical point of view, both criteria aim at quantifying how well a model $g(x|\theta)$ approximates the true, but unknown, data distribution $h(x)$. The AIC is related to an estimate of the Kullback-Leibler divergence between these two distributions, and therefore favors models that achieve a good fit without introducing unnecessary parameters. 

The DIC follows a similar idea but replaces the total number of parameters with the effective number $p$, thus reducing the penalty for parameters that are weakly constrained by the data.

Since the true distribution $h(x)$ is unknown, the absolute values of AIC and DIC are not informative on their own. For this reason, we consider only differences with respect to the best-fitting model,
\begin{equation}
\Delta X = X_i - X_{\rm \min}, \qquad X_{\rm i} \in {\mathrm{AIC}, \mathrm{DIC}},
\end{equation}
where $X_{\min}$ is the minimum value among the models under consideration. By definition, the preferred model has $\Delta X = 0$. Larger values of $\Delta X$ indicate a worse fit, with the usual interpretation: $\Delta X \geq 2$ (weak evidence), $\Delta X \geq 6$ (moderate evidence), and $\Delta X \geq 10$ (strong evidence) \cite{CosmoVerseNetwork:2025alb}.

In Tab.~\ref{tab:comparison}, we present the outcome of our analysis combining CMB, BAO, and Pantheon data with the SH0ES prior on $H_0$.

Even though the results obtained suggest EDE as the best-fit model, they do not  exclude $\Lambda_{\omega_s}$CDM and, rather, the AIC criterion provides only a moderate support in favor of EDE, while the DIC indicates only weak preference, indicating that our model remains a viable alternative to alleviate the $H_0$ tension. Nevertheless, this may be justified by the \emph{ad hoc} property of EDE to pick its effect just before recombination, while having for the $\Lambda_{\omega_s}$CDM scenario, the need of further data that constrain simultaneously $\omega_s$ and $\Omega_s$.

\begin{figure}[htb!]
\includegraphics[width=0.50\textwidth,clip]{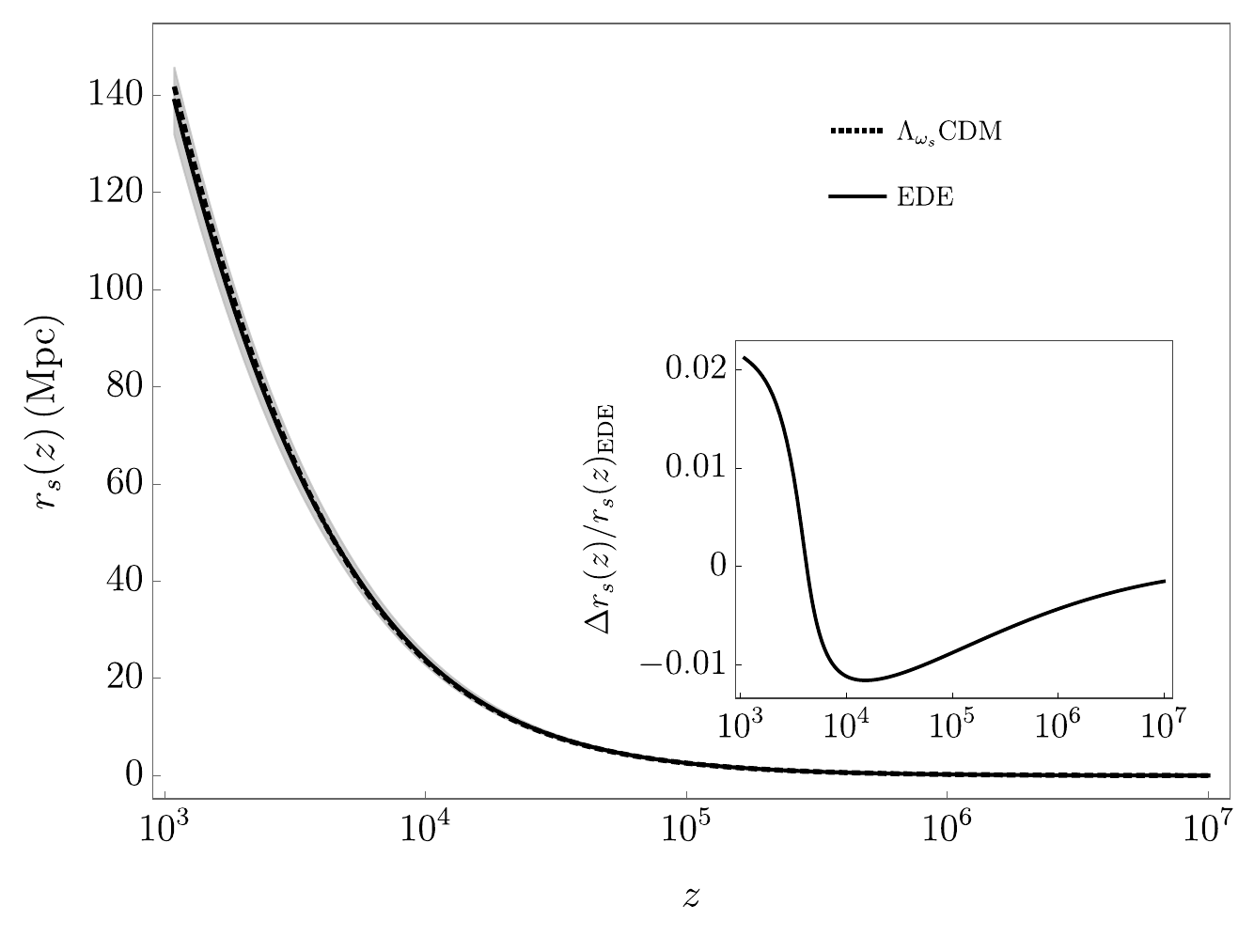}
\caption{Comoving sound horizon $r_s(z)$ for $\Lambda_{\omega_s}$CDM and EDE. 
The shaded region corresponds to a $\pm 5\%$ band around the EDE prediction. 
The inset displays the relative difference $\Delta r_s(z)/r_s(z)_{\mathrm{EDE}}$. 
The cosmological parameters are those reported in Tab.~\ref{tab:comparison}.}
\label{fig:rsede}
\end{figure}

Finally, in Fig.~\ref{fig:rsede} we show the difference between the evolution of the comoving sound horizon in $\Lambda_{\omega_s}$CDM and EDE. The result lies within the $5\%$ band around the EDE prediction, indicating only a small deviation between the two scenarios.

Summing up, our results indicate only a mild statistical preference for EDE over the $\Lambda_{\omega_s}$CDM model according to information criteria. However, this advantage is driven by the intrinsic phenomenological flexibility of the EDE framework. More precisely, EDE relies on a highly-tuned scalar field that becomes dynamically relevant only within a narrow redshift interval around recombination, effectively introducing a localized deformation of the expansion history.

In contrast, the $\Lambda_{\omega_s}$CDM scenario proposed here is based on a persistent barotropic fluid that contributes to the cosmic energy budget throughout the entire evolution. This feature reduces the model flexibility, as it cannot selectively modify the expansion rate at specific epochs. Consequently, a mildly worse statistical fit should not be interpreted as a limitation but rather as a direct consequence of the stronger physical constraints and enhanced predictivity of the model.

From a theoretical perspective, the proposed fluid admits a natural interpretation as a quasi-relativistic matter component with pressure, rather than invoking an additional scalar field with \emph{ad hoc} dynamics. Furthermore, the model represents a minimal extension of the $\Lambda$CDM paradigm, introducing a single additional parameter that remains testable across different cosmological epochs. This makes the scenario falsifiable with future observations and capable of producing correlated signatures at multiple redshifts, in contrast to the localized nature of EDE.

\section{Physical interpretation of our multi-fluid cosmological scenario}\label{sec7}

A central aspect of the $\Lambda_{\omega_s}$CDM paradigm concerns the physical interpretation of the additional barotropic component introduced in Eq. \eqref{eq:rhos_standard}. In particular, beyond its phenomenological success in alleviating the $H_0$ tension, one needs to clarify, 

\begin{itemize}
    \item[-] the physical origin of this fluid, 
    \item[-] why such a fluid can exist as a stable and persistent constituent of the cosmic energy budget. 
\end{itemize}

Two main scenarios are discussed below. In the first part, we explore the origins of our additional component, remarking why it is expected not to decay, whereas in the second part, we exclude its origin in terms of matter and/or radiation deformations. 

\subsection{Physical origin and stability of the barotropic fluid}

Our additional component is characterized by a constant barotropic equation of state, that in view of the above numerical findings, approximately provides a density,
\begin{equation}
\rho_s(a) \simeq 10^{-5}\,a^{-3.87}\,.
\end{equation}
This scaling law immediately implies that the fluid behaves as an intermediate component between radiation and pressureless matter, corresponding to a quasi-relativistic species.

From a physical perspective, the stability of this fluid can be understood as a consequence of three fundamental properties:

\paragraph*{(i) Absence of internal instabilities.}
The assumption of a constant equation of state ensures that the fluid admits a well-defined adiabatic sound speed,
\begin{equation}
c_s^2 = \frac{dP_s}{d\rho_s} = \omega_s > 0\,,
\end{equation}
guaranteeing the absence of gradient instabilities. Unlike scalar field realizations with time-dependent potentials, no dynamical evolution of $\omega_s$ is required and the fluid corresponds to a stable thermodynamic phase.

\paragraph*{(ii) Cosmological attractor behavior.}
The scaling $\rho_s \propto a^{-3(1+\omega_s)}$ represents a self-consistent solution of the continuity equation in an expanding background. Accordingly, it defines a cosmological attractor: once produced in the early Universe, the fluid persists without requiring fine-tuned initial conditions or triggering mechanisms localized in time. This is in net contrast with EDE scenarios, where the additional component becomes relevant only in a narrow redshift window and may appear more probable. 

\paragraph*{(iii) Decoupling and absence of decay channels.}
In our framework, the additional fluid is assumed to be non-interacting with the standard components, except through gravity. This implies that no efficient decay channels into radiation or matter are available. Consequently, the fluid is not transient, but rather a relic component that survives throughout the entire cosmic evolution. The subdominance at late times naturally explains why it has not been directly detected so far. Cosmic bounds seem to confirm this theoretical scheme.

\paragraph*{(iv) Non-clustering nature.}
As shown above, perturbations in the fluid do not develop a growing mode and therefore do not contribute to structure formation. This behavior is analogous to that of relativistic species, indicating that the fluid acts as a smooth background component rather than a clustering one. This property further stabilizes the fluid against gravitational collapse and prevents fragmentation.

From a microscopic viewpoint, several interpretations can be therefore proposed for our new fluid, corresponding to:
\begin{itemize}
\item[-] a relic quasi-relativistic particle species that decoupled in the early Universe,
\item[-] an effective description of microscopic degrees of freedom beyond the Standard Model of particle physics,
\item[-] a remnant of a more fundamental high-energy sector, whose macroscopic behavior is incorporated in our effective barotropic factor.
\end{itemize}

\subsection{Justifying matter with pressure nature}

The interpretation of the additional component as ``matter with pressure'' can be physically concrete by identifying it with a collisionless particle species characterized by a non-negligible velocity dispersion. In this case, the effective equation of state arises from kinetic theory as
\begin{equation}
\omega_s = \frac{P_s}{\rho_s} = \frac{1}{3}\langle v^2 \rangle \,,
\end{equation}
where $\langle v^2 \rangle$ denotes the mean squared velocity of the particles.

For $\omega_s \sim 0.3$, the fluid corresponds to a quasi-relativistic component, interpolating between radiation and pressureless matter. Unlike cold dark matter, the presence of pressure prevents the growth of perturbations below the corresponding Jeans scale, thereby explaining the absence of clustering found in our analysis. At the same time, the fluid contributes to the background expansion, modifying the early-time dynamics without significantly affecting structure formation.

This picture naturally emerges in scenarios involving relic particle species that decouple while still semi-relativistic, or from hidden sectors interacting only gravitationally with the Standard Model. In this sense, the additional barotropic fluid should be regarded as a stable, non-interacting, quasi-relativistic matter component, rather than as an effective deformation of known cosmological fluids.

Accordingly, we provide our final interpretation for our fluid, namely \emph{the cosmic medium is composed of a standard clustering cold dark matter sector plus an additional non-clustering matter with pressure sector, the latter acting as an intermediate quasi-relativistic background component}, whose main differences with respect to the standard case are listed in Tab. \ref{tab:comparison_fluid}.

\begin{table*}[t]
\centering
\small
\begin{tabular}{l c c c c}
\hline\hline
Component & $\omega_i$ & $\rho(a)$ & Clustering & Role \\
\hline\hline
Cold dark matter & $0$ & $a^{-3}$ & Yes & Structure formation \\
Radiation & $1/3$ & $a^{-4}$ & No & Early Universe expansion \\
Matter with pressure & $\lesssim 0.3$ & $a^{-3(1+\omega_s)}$ & Suppressed & Background expansion \\
\hline\hline
\end{tabular}
\caption{Comparison between cold dark matter, radiation, and the additional barotropic component in the $\Lambda_{\omega_s}$CDM scenario. The parameter $\omega_i$ denotes the equation of state of each component, with $\omega_m=0$ for cold dark matter and $\omega_r=1/3$ for radiation. Cold dark matter is pressureless and clusters efficiently, driving structure formation. Radiation is fully relativistic, does not cluster, and dominates the early-time expansion. The additional component is characterized by a constant positive $\omega_s$, lying between these two limits. It behaves as a quasi-relativistic sector that modifies the expansion history, while its pressure suppresses the growth of perturbations.}
\label{tab:comparison_fluid}
\end{table*}

Taken together, these properties suggest that the additional barotropic component should not be interpreted as an effective deformation of known fluids, but rather as a genuinely independent sector of the cosmological energy budget. Hence, in the following subsection, we explicitly demonstrate that the additional component introduced in the $\Lambda_{\omega_s}$CDM model cannot be reabsorbed into a deformation of the standard matter or radiation sectors. Instead, it must be interpreted as a genuinely new fluid, characterized by independent physical properties and a distinct scaling behavior.

\subsection{Beyond deforming matter and radiation}

A natural question is whether the additional barotropic contribution discussed above can be reinterpreted as a small deformation of the standard matter or radiation sectors, rather than as an independent fluid. 

Accordingly, one may wonder whether it is sufficient to modify the usual scalings according to
\begin{equation}
\rho_m(a)=\rho_{m}a^{-3-\epsilon_m},
\qquad
\rho_r(a)=\rho_{r}a^{-4-\epsilon_r},
\label{eq:deformed_matter_radiation}
\end{equation}
with $|\epsilon_m|\ll 1$ and $|\epsilon_r|\ll 1$. At first sight, this possibility may appear appealing, since it avoids the introduction of one more cosmological constituent. However, a closer investigation shows that the two pictures are not equivalent. 

Moreover, the description in terms of an additional barotropic fluid is theoretically more robust and likely, from a  phenomenological viewpoint, more appropriate for the mechanism required to alleviate the Hubble tension.

To clarify this point and compare the hypothesis in Eq. \eqref{eq:deformed_matter_radiation} with Eq. \eqref{eq:rhos_standard}, it is useful to distinguish three main levels, as below report.
\begin{enumerate}
\item[(i)] An additional fluid, characterized by its own conservation law and equation of state.
\item[(ii)] A deformation of matter or radiation viewed as an exact modified scaling law.
\item[(iii)] A first-order logarithmic expansion of the deformed scaling in the limit of small $\epsilon$.
\end{enumerate}
We intend to show below that the first option naturally leads to a viable early-time mechanism to exploit cosmic tension.

The standard background evolution within the $\Lambda$CDM model is given by
$E^2(a)=
\Omega_{m}a^{-3}
+
\Omega_{r}a^{-4}
+
\Omega_{\Lambda}$
and, by virtue of Eq.~\eqref{eq:rhos_standard}, it can be extended to the $\Lambda_{\omega_s}$CDM scenario as in Eq.~\eqref{eq:hubble}, ensuring that the new contribution is \emph{additive}.

This means that it corresponds to a separate physical species, determined by its own present abundance $\Omega_{s0}$ and with a definite redshift scaling fixed by the barotropic factor $\omega_s$. 

In particular, the continuity equation for the new fluid only appears
\begin{equation}
\dot{\rho}_s(t)+3H(t)(1+\omega_s)\rho_s(t)=0
\label{eq:cont_s}
\end{equation}
immediately suggesting unambiguously the interpretation of a \emph{multi-fluid cosmology}.

Conversely, if one replaces matter or radiation with a deformed law such as Eq.~\eqref{eq:deformed_matter_radiation}, the corresponding term in the Hubble function becomes \emph{multiplicative}, implying 

\begin{subequations}
    \begin{align}
&E^2(a)
=
\Omega_{m}a^{-3-\epsilon_m}
+
\Omega_{r}a^{-4}
+
\Omega_{\Lambda},\label{eq:E2_def_m}\\
&        
E^2(a)
=
\Omega_{m}a^{-3}
+
\Omega_{r}a^{-4-\epsilon_r}
+
\Omega_{\Lambda}.\label{eq:E2_def_r}
    \end{align}
\end{subequations}
This is conceptually different from adding a new species; in fact, in Eqs.~\eqref{eq:E2_def_m} and \eqref{eq:E2_def_r} the standard fluids themselves no longer obey their usual conservation laws. 

In turn, unless one is willing to interpret these expressions as phenomenological fitting functions, one must explain why matter or radiation is not conserved separately.

Indeed, taking the matter- and radiation-deformed cases, respectively as,
\begin{subequations}
\begin{align}
&\rho_m(a)=\rho_{m}a^{-3-\epsilon_m}\,\Rightarrow \dot{\rho}_m(t)+3H(t)\rho_m(t)=-\epsilon_m H(t)\rho_m(t),\label{eq:rhom_def}\\
&\rho_r(a)=\rho_{r}a^{-4-\epsilon_r}\,\Rightarrow \dot{\rho}_r(t)+4H(t)\rho_r(t)=-\epsilon_r H(t)\rho_r(t).\label{eq:rhor_def}
\end{align}
\end{subequations}
Thus, a nonzero $\epsilon_m\neq0$ or $\epsilon_r\neq0$ implies that matter or radiation are not conserved in the standard sense.

Eqs.~\eqref{eq:rhom_def} and \eqref{eq:rhor_def} require either an interaction with another hidden sector, a source term induced by modified gravity, particle production, or some nontrivial coarse-grained effective description. In the absence of an underlying microphysical model, the deformation parameters $\epsilon_m$ and $\epsilon_r$ do not describe a fundamental cosmological component but rather encode a violation of the standard conservation laws of the known fluids.

Our additional fluid picture is therefore conceptually and theoretically preferable, since \emph{it preserves the standard matter and radiation sectors and modifies the cosmological evolution only through one extra, explicitly defined barotropic contribution}. 

In this sense, the extension in Eq.~\eqref{eq:hubble} is minimal, whereas the deformations in Eqs.~\eqref{eq:E2_def_m} and \eqref{eq:E2_def_r} are minimal only at the level of notation, albeit physically more complicated.

To stress the difference, the effective exponents can be evaluated.  Our best fits favor a quasi-relativistic component, thus having, 
\begin{equation}
3(1+\omega_s)\simeq 3(1+0.29)\simeq 3.87\,\Rightarrow \epsilon_m=3\omega_s\simeq 0.87.
\label{eq:best_fit_exponent}
\end{equation}
This value is manifestly not small, implying that \emph{the additional fluid cannot be consistently reabsorbed into a small deformation of dust matter}. 

If instead one attempts to identify the new component with deformed radiation, one obtains
\begin{equation}
4+\epsilon_r=3(1+\omega_s)\,\Rightarrow\epsilon_r=3\omega_s-1\simeq -0.13. 
\end{equation}
Numerically, this is smaller in magnitude than $\epsilon_m$, and therefore, radiation deformation may look more plausible. However, this outcome turns out to be unphysical and/or disfavored for at least four main conceptual reasons, as reported below.

\begin{itemize}
    \item[-] First, even if $|\epsilon_r|$ is formally below unity, it is not small enough to control logarithmic expansion over the redshift interval relevant to recombination. Expanding Eq.~\eqref{eq:deformed_matter_radiation} to first order implies that the perturbative expansion remains valid only if $
|\epsilon \ln a|\ll 1$.
At recombination, $a_\ast\sim 10^{-3}$, so that $|\ln a|\approx 7$. Hence, for the radiation-like fluid one finds
\begin{equation}
|\epsilon_r \ln a_\ast|
\approx
0.13\times 7
\approx
0.9,
\label{eq:epsr_fail}
\end{equation}
which is not a small quantity. Therefore, the first-order treatment already breaks down in the regime where the model is supposed to operate. The same conclusion holds \emph{a fortiori} for the matter-like case, for which
\begin{equation}
|\epsilon_m \ln a_\ast|\approx 0.87\times 7\approx 6.
\end{equation}
Thus, the logarithmic approximation cannot reproduce the physics of our additional fluid across the redshifts relevant to the sound horizon problem.

\item[-] Second, the deformation picture modifies the standard fluids and radiation is is linked to the thermal plasma, to the relativistic degrees of freedom, to the recombination history and to the structure of the acoustic peaks. Accordingly, modifying its dilution law is not equivalent to adding a new quasi-relativistic component. The former changes the identity of the radiation bath itself, whereas the latter changes the total energy budget while leaving the standard sectors conceptually intact, that is what we intended to present adding the fluid in Eq. \eqref{eq:rhos_standard}.

\item[-] Third, the matter and radiation deformed fluids provide
\begin{subequations}
\begin{align}
\delta E^2_m(a) &\simeq -\Omega_{m}\,a^{-3}\,\epsilon_m \ln a,\label{a11}\\
\delta E^2_r(a) &\simeq -\Omega_{r}\,a^{-4}\,\epsilon_r \ln a,\label{a12}
\end{align}
\end{subequations}
showing that the corrections are not controlled by an independent abundance, but to $\Omega_{m}$ or $\Omega_{r}$. 

This fact is experimentally and theoretically quite restrictive since it needs to modify the two well-constrained abundances of dust and photons.

\item[-] Fourth, the additive form in Eq.~\eqref{eq:hubble} is also much more stable as it remains positive definite for $\Omega_{s}>0$ and any real $\omega_s$. On the other hand, the logarithmic expressions in Eqs.~\eqref{a11} and \eqref{a12} can become pathological outside the narrow interval where the expansion is valid. For instance, if $\epsilon_r < 0$, the bracket $1 - \epsilon_r \ln a$ increases as $a$ decreases and may even become negative at sufficiently small scale factors, signaling the breakdown of the perturbative description.

\item[-] Last but not least, another key aspect concerns perturbations. The model developed in this work, as expected theoretically, treats the additional component as a nonclustering or weakly clustering species, whose main effect is on the background expansion and on the sound horizon, while preserving the standard behavior of the growth sector to a very good approximation. This interpretation is consistent because the fluid is introduced as a separate constituent. In contrast, deforming matter directly raises an immediate conceptual problem: \emph{if the matter sector itself acquires a modified dilution law, should it still cluster as standard cold matter}? If yes, one must explain why its background scaling changes while its perturbative behavior remains unchanged. If not, the standard matter component breaks down. 

Analogously, since radiation is tightly connected to the photon-baryon plasma and to the pre-recombination sound speed, modifying its redshift dependence necessarily affects several aspects of the early-Universe dynamics simultaneously. By introducing instead one more barotropic fluid, the model isolates the new physics into an additional sector whose impact can be consistently tracked both in the background evolution and in the pre-recombination plasma.

\end{itemize}


\section{Conclusions and perspectives}\label{Sec.8}

In this work, motivated by the $H_0$ tension, we explored an extension of our present cosmological paradigm by \emph{introducing an additional barotropic fluid}, interpreted as a matter-like component, exhibiting non-zero pressure. 

Our corresponding paradigm, dubbed the $\Lambda_{\omega_s}$CDM scenario, consists of the presence of this extra component, modifying the cosmological dynamics at all scales. More precisely, our fluid shows a barotropic factor that relies inside the Zeldovich limit, quite different from any dark component. To explore its effects in cosmology, we provided particular emphasis on the early-time expansion history and, thus, reformulating the Friedmann equations. 

In this respect, we showed that the inclusion of this fluid altered the total energy density and the Hubble expansion rate, leading to a modified cosmological evolution characterized by a four-component Universe composed of radiation, dust, dark energy and our fluid, under the form of a quasi-relativistic matter constituent.

We demonstrated that the new fluid naturally contributed to the early-time thermal bath, effectively yielding a three-fluid description of the primordial Universe. Within this framework, we found that the additional component enhanced the pre-recombination expansion rate and modified the sound speed of the photon-baryon plasma. As a direct consequence, the comoving sound horizon decreased, providing a physical mechanism to increase the inferred value of the Hubble constant.

We then analyzed the analytical behavior of the scale factor in the relevant asymptotic regimes. For the radiation-fluid system, we derived the implicit solution in terms of hypergeometric functions and identified the transition at $\omega_s = 1/3$ as the critical threshold separating distinct dynamical behaviors. For $\omega_s < 1/3$, we found that radiation dominated the asymptotic past, with the additional fluid providing subleading corrections that preserved the standard radiation era. Conversely, for $\omega_s > 1/3$, we showed that the extra component dominated the early-time expansion and replaced radiation as the leading contribution. We obtained analogous results in the matter-fluid system, where the new component modified the early-time dynamics for any $\omega_s > 0$, while the standard dust-dominated behavior was recovered at late times. 

We further quantified these effects by comparing the numerical solution of the full three-fluid system with the analytical two-fluid $\Lambda$CDM solution. We found that for $\omega_s < 1/3$ the deviation remained negligible throughout the early-time range, confirming that the model preserved the standard radiation-dominated epoch. In contrast, for $\omega_s > 1/3$, we observed significant deviations already at very early-times, indicating that the additional fluid controlled the expansion history in this regime. This demonstrated that the model provided a direct modification of the early-Universe dynamics rather than a perturbative correction to $\Lambda$CDM, suggesting the existence of a purely novel species whose origin would be responsible for cosmic tensions and possibly for additional observable effects.

At the perturbative level, we thus analyzed the evolution of linear matter fluctuations by deriving the equations that govern $\delta(a)$, the growth factor $D(a)$, the growth rate $f(a)$ and the observable $f\sigma_8(a)$. In particular, we wondered whether the fluid significantly influences linear cosmological perturbations. Similarly to radiation, the additional fluid does not significantly cluster and, accordingly, we showed that it did not contribute to the source term driving structure formation. Although it modified the background expansion, its impact on the perturbative sector remained subdominant. In particular, we found that deviations from $\Lambda$CDM predictions for all relevant observables remained within a few percent over the entire evolution, thereby preserving the successful description of LSS.

We performed a Bayesian analysis of the model parameters by implementing the $\Lambda_{\omega_s}$CDM scenario in a modified version of the CLASS Boltzmann code and exploring the parameter space with MontePython. We considered a hierarchical combination of cosmological probes, including \textit{Planck} 2018 CMB data, DESI DR2 BAO measurements, Pantheon SNe Ia, observational Hubble data, and the SH0ES prior on $H_0$. This allowed us to assess the statistical relevance of the additional fluid under different observational conditions.

We found that, in the absence of the SH0ES prior, the parameters $\omega_s$ and $\Omega_s$ were driven toward the lower bounds of their priors and the model effectively reduced to the $\Lambda$CDM scenario. This indicated that early-time probes alone did not require the presence of the additional component. When the SH0ES prior was included, the degeneracy with $H_0$ was lifted and the extra fluid emerged as a nonvanishing contribution. We obtained positive and stable mean values for both $\omega_s$ and $\Omega_s$, with consistent results across different dataset combinations. This showed that the additional barotropic fluid became relevant precisely when the tension on the Hubble constant is taken into account.

These results established that the extra fluid acted as a physical mechanism capable of modifying the pre-recombination expansion rate, reducing the sound horizon and shifting the inferred value of $H_0$ toward higher values. 

Very precisely, our novel fluid differs conceptually and dynamically from EDE since, in standard EDE scenarios, the additional contribution is modeled by an axion-like scalar field with a finely tuned potential, which remains frozen at early-times due to Hubble friction and becomes dynamically relevant only within a narrow redshift window around recombination. Conversely, the fluid considered here is characterized by a constant barotropic equation of state that can be justified to remain stable throughout the entire cosmological evolution. As a consequence, \emph{it does not require any triggering mechanism or phase transition to become active, but instead provides a continuous, albeit subdominant, contribution to the total energy budget}.

From a physical perspective, this difference implies that, while EDE introduces a transient and highly dynamical modification of the expansion history, the barotropic fluid yields a smoother and more persistent correction to the background evolution, scaling as $\rho_s\propto a^{-3(1+\omega_s)}$. Moreover, EDE relies on additional scalar degrees of freedom and associated model parameters (e.g., the field mass, decay constant, and initial value of the scalar field), whereas the present scenario can be interpreted as an effective description of a quasi-relativistic component without invoking explicit microphysical representations. 


These outcomes suggested why at late-times the fluid seems non-influent. Indeed, in this regime, when a Gaussian prior was imposed, the inferred parameter values became consistent with those obtained from the full dataset combination. This confirmed that the primary physical effects of the new component were associated with the early-time expansion.

Overall, our analysis demonstrated that the introduction of an additional barotropic fluid provided a consistent and minimal extension of our current cosmological understanding, capable of modifying the early-Universe dynamics without changing both BBN and the growth of cosmic structures. Hence, our model remained compatible with current observations and \emph{offered a competitive alternative to existing early-time solutions to cosmological tensions with one or multiple axion-like fields}. 

In future works, we will investigate the microphysical origin of the additional barotropic sector and assess whether it can emerge from fundamental frameworks, such as effective field theories, quasi-relativistic relics, or nonstandard thermal histories. We will extend the perturbation analysis beyond the nonclustering approximation and test the model against a broader set of LSS observables. We will also perform a systematic comparison with alternative early-time scenarios, including different realizations of EDE and interacting models, in order to quantify the relative statistical performance and physical robustness of the $\Lambda_{\omega_s}$CDM framework. Last but not least, we will explore better possible connections with other cosmological anomalies, including the $S_8$ tension and deviations from standard recombination, to determine whether the presence of an additional barotropic fluid can provide a unified description of multiple observational discrepancies. Finally, we intend to explore the impact of our new barotropic fluid at very early-times, understanding how the radiation-dominated hot plasma would change according to the fact that the extra fluid dominates over matter at that period. 

\section*{Acknowledgments}
\noindent YC and OL acknowledge financial support from Roberto Della Ceca and the Brera National Institute for Astrophysics. The authors are also very grateful to Vivian Poulin for debates and discussions on the cosmological tensions. Numerical results obtained in this paper were computed through resources from Universe and Particles Laboratory of Montpellier (LUPM). The authors thank LUPM for providing the technical support, computing and storage facilities.

\bibliography{biblio}

\appendix

\section{MCMC contour plots for the $\Lambda_{\omega_s}$CDM model}\label{AppA}

In this Appendix, we present the MCMC contour plots obtained from the various analyses, which combine both early- and late-time cosmological probes in order to explore the parameter space of the model. The different dataset combinations are systematically organized based on the inclusion or exclusion of CMB measurements. In particular, we consider configurations that include CMB data and Pantheon SNe Ia, either with or without the SH0ES prior, as well as scenarios in which CMB information is entirely excluded; in the latter case, we further distinguish between analyses performed with or without the inclusion of OHD data. For the subset of analyses relying exclusively on late-time probes, we also examine the effect of imposing a Gaussian prior on $\Omega_s$. This additional assumption is motivated by the fact that, in the absence of early-time constraints, $\Omega_s$ remains largely unconstrained when flat priors are adopted, thus requiring a more informative prior to meaningfully restrict its parameter space.

\begin{figure*}[htb!]
\includegraphics[width=\hsize,clip]{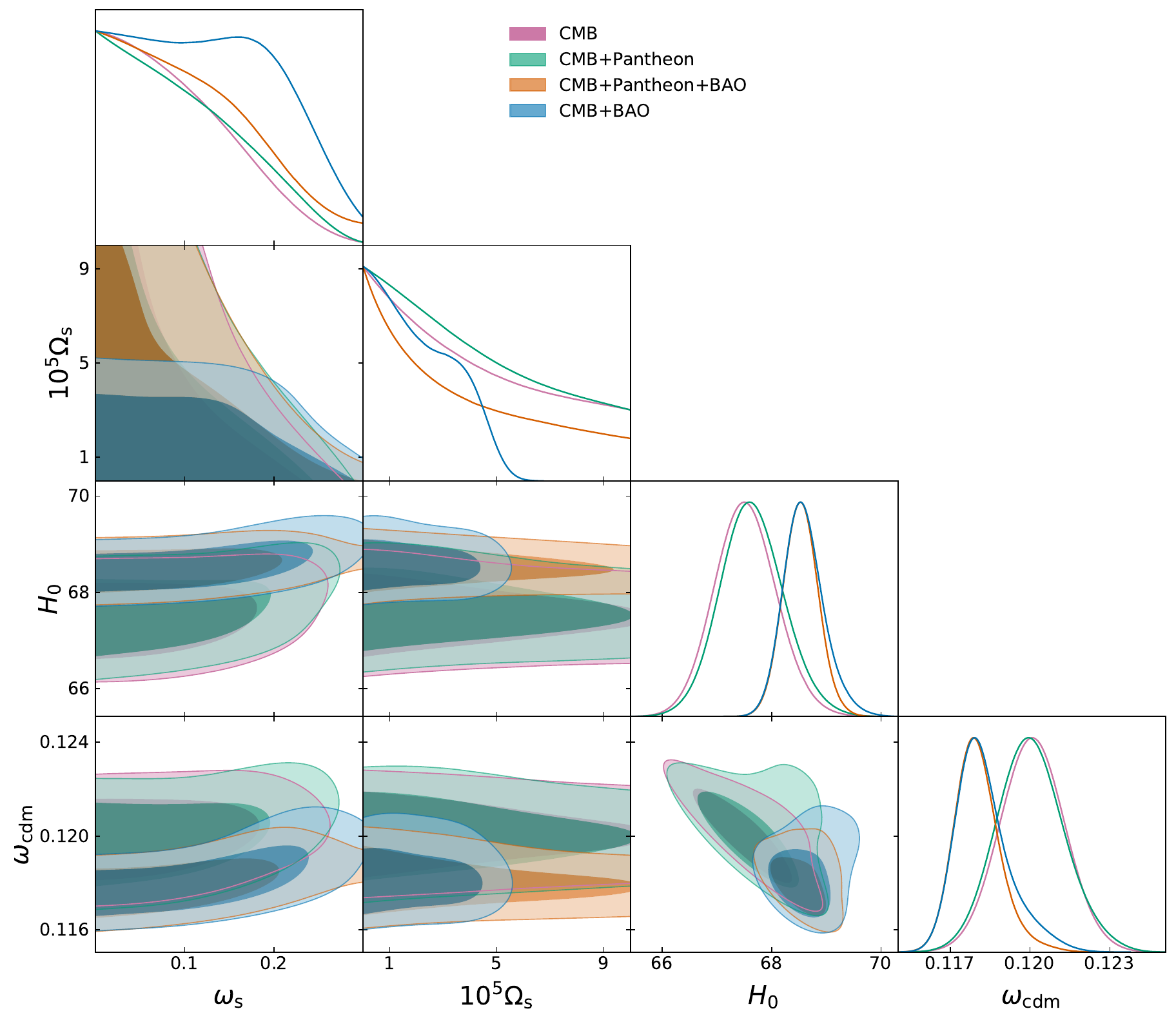}
\caption{Posterior distributions for the $\Lambda_{\omega_s}$CDM model without the SH0ES prior. The combinations of cosmological probes considered are: \textit{Planck} 2018 CMB, \textit{Planck} 2018 CMB + BAO (DESI DR2), \textit{Planck} 2018 CMB + Pantheon, and \textit{Planck} 2018 CMB + BAO (DESI DR2) + Pantheon.}
\label{fig:nosh0es}
\end{figure*}

\begin{figure*}[htb!]
\includegraphics[width=\hsize,clip]{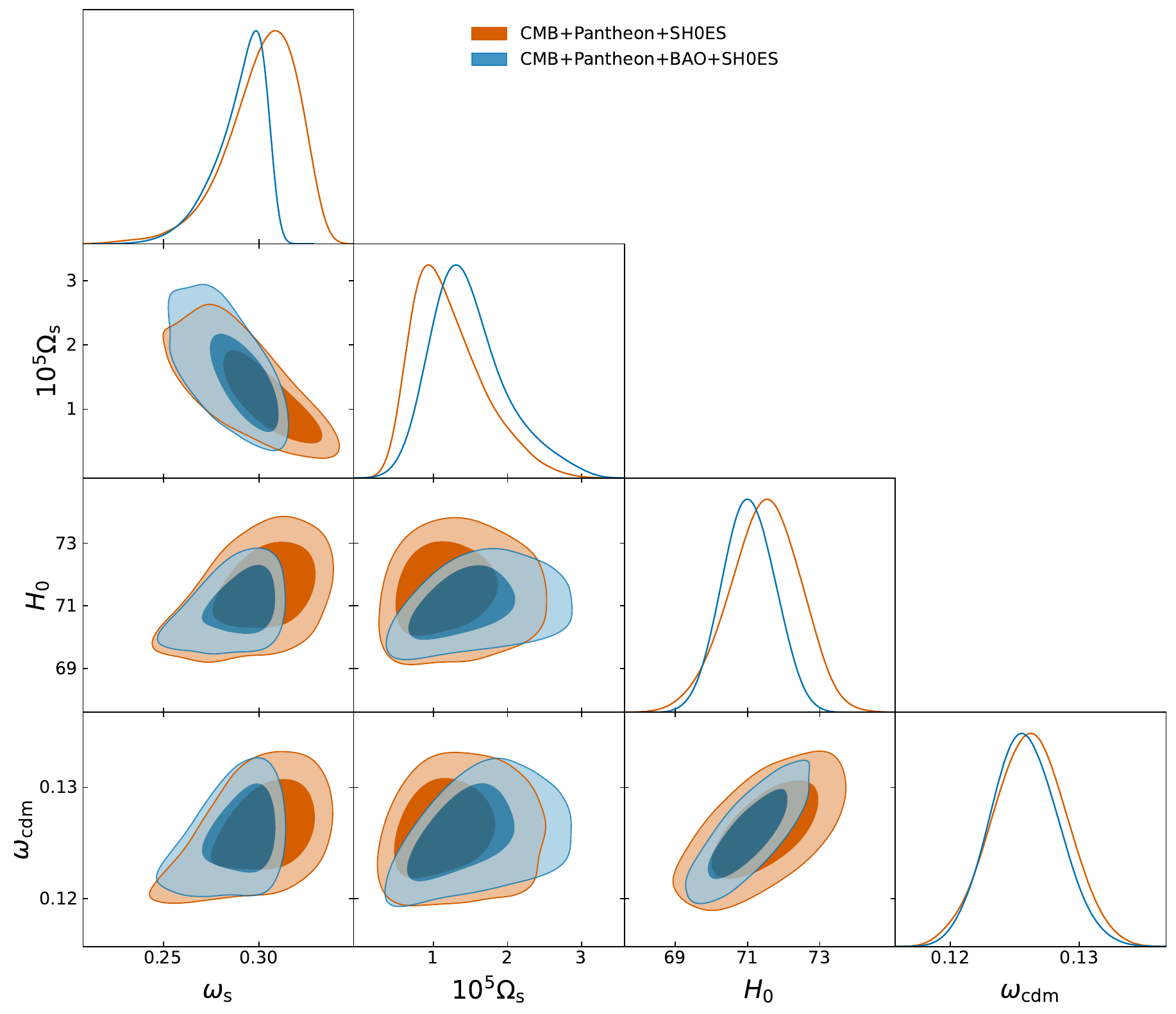}
\caption{Posterior distributions for the $\Lambda_{\omega_s}$CDM model including the SH0ES prior. The combinations of cosmological probes considered are: \textit{Planck} 2018 CMB + BAO (DESI DR2) + Pantheon + SH0ES, and \textit{Planck} 2018 CMB + Pantheon + SH0ES.}
\label{fig:sh0es}
\end{figure*}

\begin{figure*}[htb!]
\includegraphics[width=\hsize,clip]{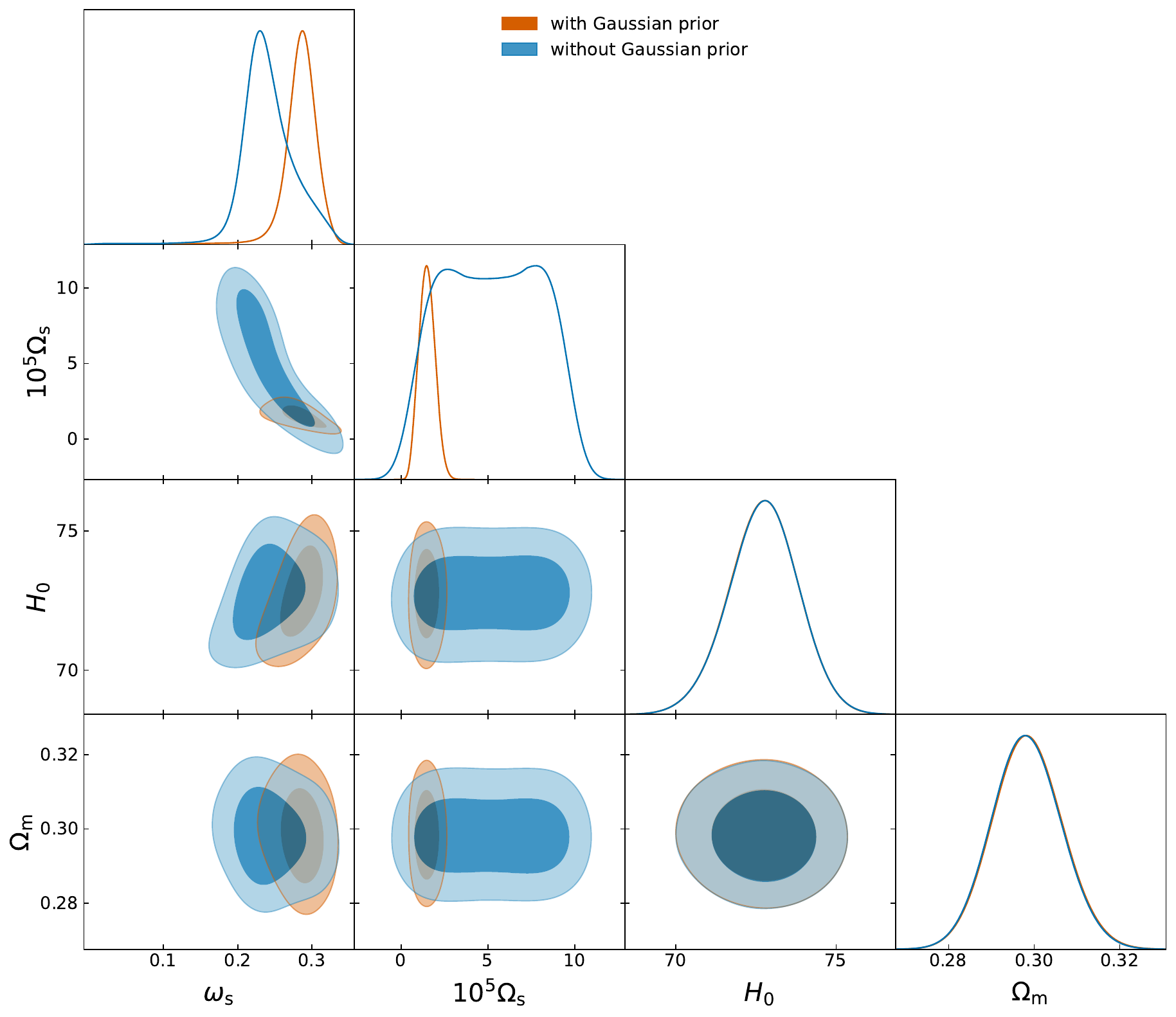}
\caption{Posterior distributions for the $\Lambda_{\omega_s}$CDM model using only late-time probes, excluding CMB data. The combinations of cosmological probes considered are BAO (DESI DR2) + Pantheon + SH0ES, with and without a Gaussian prior on $\Omega_s$.}
\label{fig:nocmb}
\end{figure*}

\begin{figure*}[htb!]
\includegraphics[width=\hsize,clip]{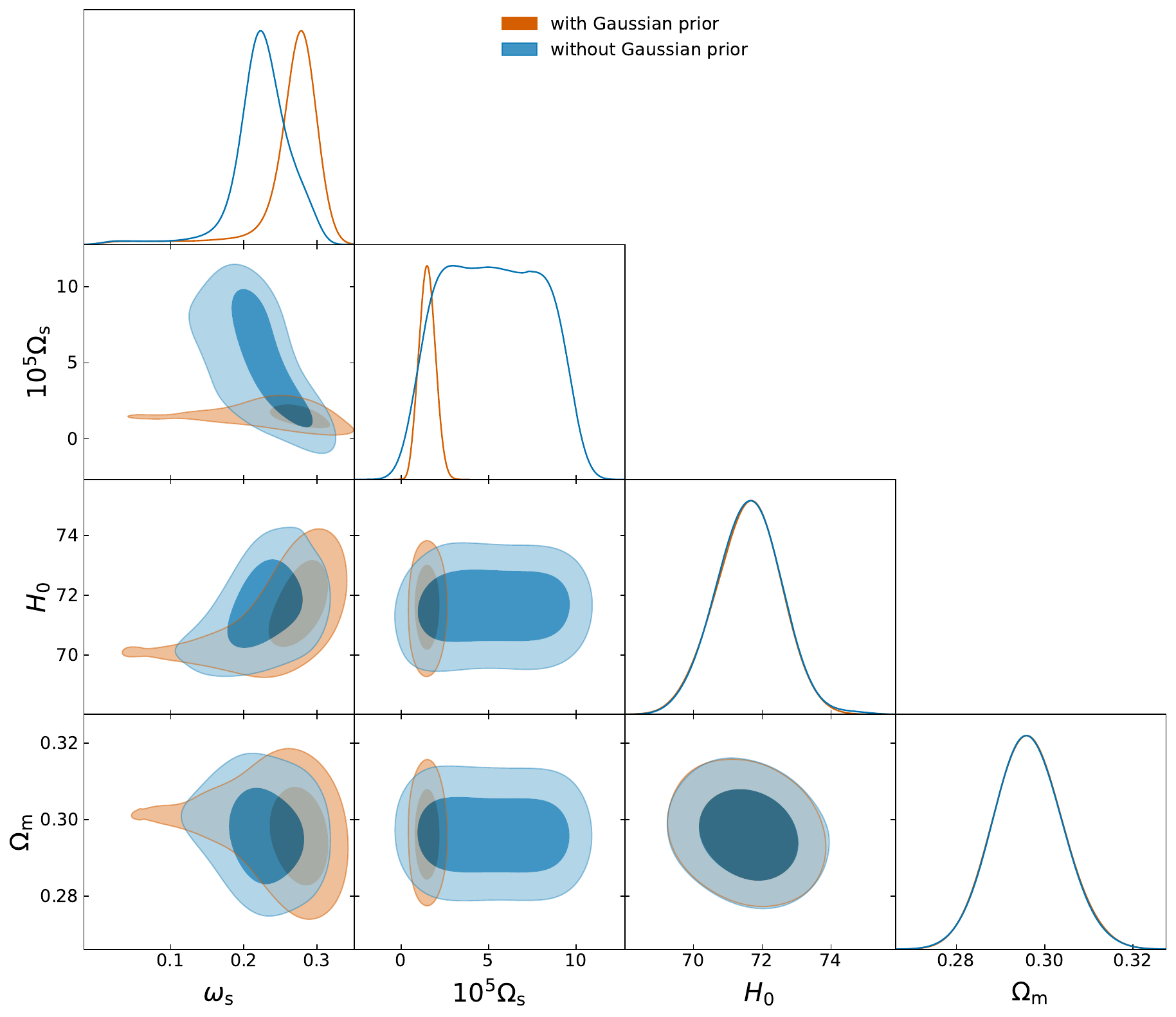}
\caption{Posterior distributions for the $\Lambda_{\omega_s}$CDM model using only late-time probes, excluding CMB data. The combinations of cosmological probes considered are BAO (DESI DR2) + Pantheon + OHD + SH0ES, with and without a Gaussian prior on $\Omega_s$.}
\label{fig:nocmbOHD}
\end{figure*}

\end{document}